\documentclass[twocolumn,aps,superscriptaddress,showpacs,showkeys,amsmath,amssymb,floatfix, longbibliography]{revtex4-1}
\usepackage[colorlinks=true,citecolor=blue,filecolor=blue,linkcolor=blue,urlcolor=blue,pdftex]{hyperref}
\usepackage{currfile} 
\usepackage{graphicx}
\usepackage{dcolumn}
\usepackage{bm}
\usepackage[section]{placeins} 

\usepackage{isotope} 
\usepackage{upgreek} 
\usepackage{cancel}
\usepackage{soul} 
\usepackage[separate-uncertainty=true]{siunitx} 

\usepackage{color}
\usepackage{amsmath}
\usepackage{multirow}
\usepackage{xspace}
\usepackage{orcidlink} 

\newcommand{\radon}{\ensuremath{\isotope[222]{Rn}}\xspace}

\newcommand{\cevns}{CE$\nu$NS\xspace}

\newcommand{\gxepur}{GXe-PUR\xspace}
\newcommand{\lxepur}{LXe-PUR\xspace}
\newcommand{\cry}{CRY\xspace}
\newcommand{\tpc}{TPC\xspace}
\newcommand{\rrs}{RRS\xspace}

\newcommand{\noRRS}{\textit{No RRS}\xspace}
\newcommand{\GXeOnlyRRS}{\textit{GXe-only RRS}\xspace}
\newcommand{\GXeLXeRRS}{\textit{GXe+LXe RRS}\xspace}

\newcommand{\RadonValueExpectation}{4.2$_{-0.7}^{+0.5}$\,\si{\micro \becquerel \per \kilo \gram}\xspace}
\newcommand{\RadonValuePreSRZero}{(3.62$\,\pm\,$0.07\,stat.$\,\pm\,$0.17 sys.)\,\si{\micro \becquerel \per \kilo \gram}\xspace}
\newcommand{\RadonValueSRZero}{(1.87$\,\pm\,$0.02\,stat.$\,\pm\,$0.09 sys.)\,\si{\micro \becquerel \per \kilo \gram}\xspace}
\newcommand{\RadonValueSROne}{(0.90$\,\pm\,$0.01\,stat.$\,\pm\,$0.07 sys.)\,\si{\micro \becquerel \per \kilo \gram}\xspace}
\newcommand{\RadonReductionFactorSRZero}{(\num{1.94(4)})\xspace}
\newcommand{\RadonReductionFactorSROne}{(\num{4.01(28)})\xspace}
\newcommand{\RadonHalfLife}{\SI{3.8}{\day}\xspace}
\newcommand{\RadonValueDesign}{\SI{1}{\micro \becquerel \per \kilo \gram}\xspace}
\newcommand{\RadonValueOneTonMain}{\SI{13.3(5)}{\micro \becquerel \per \kilo \gram}\xspace}
\newcommand{\RadonValueOneTonLow}{\SI{4.5(1)}{\micro \becquerel \per \kilo \gram}\xspace}
\newcommand{\RadonReductionWrtOneTon}{\num{15}\xspace}
\newcommand{\RadonReductionWrtLZ}{\num{5}\xspace}
\newcommand{\ExtractionFitVal}{\num{0.88(6)}\xspace}
\newcommand{\RRRSFitVal}{\num{187}\xspace}

\usepackage{hyperref}
\usepackage[mathlines]{lineno}
\newcommand*\patchAmsMathEnvironmentForLineno[1]{%
  \expandafter\let\csname old#1\expandafter\endcsname\csname #1\endcsname
  \expandafter\let\csname oldend#1\expandafter\endcsname\csname end#1\endcsname
  \renewenvironment{#1}%
     {\linenomath\csname old#1\endcsname}%
     {\csname oldend#1\endcsname\endlinenomath}}%
\newcommand*\patchBothAmsMathEnvironmentsForLineno[1]{%
  \patchAmsMathEnvironmentForLineno{#1}%
  \patchAmsMathEnvironmentForLineno{#1*}}%
\AtBeginDocument{%
\patchBothAmsMathEnvironmentsForLineno{equation}%
\patchBothAmsMathEnvironmentsForLineno{align}%
\patchBothAmsMathEnvironmentsForLineno{flalign}%
\patchBothAmsMathEnvironmentsForLineno{alignat}%
\patchBothAmsMathEnvironmentsForLineno{gather}%
\patchBothAmsMathEnvironmentsForLineno{multline}%
}

\begin{document}


\title[Radon Removal in XENONnT down to the Solar Neutrino Level]{Radon Removal in XENONnT down to the Solar Neutrino Level}



\newcommand{\bologna}{\affiliation{Department of Physics and Astronomy, University of Bologna and INFN-Bologna, 40126 Bologna, Italy}}
\newcommand{\chicago}{\affiliation{Department of Physics, Enrico Fermi Institute \& Kavli Institute for Cosmological Physics, University of Chicago, Chicago, IL 60637, USA}}
\newcommand{\coimbra}{\affiliation{LIBPhys, Department of Physics, University of Coimbra, 3004-516 Coimbra, Portugal}}
\newcommand{\columbia}{\affiliation{Physics Department, Columbia University, New York, NY 10027, USA}}
\newcommand{\lngs}{\affiliation{INFN-Laboratori Nazionali del Gran Sasso and Gran Sasso Science Institute, 67100 L'Aquila, Italy}}
\newcommand{\mainz}{\affiliation{Institut f\"ur Physik \& Exzellenzcluster PRISMA$^{+}$, Johannes Gutenberg-Universit\"at Mainz, 55099 Mainz, Germany}}
\newcommand{\mpik}{\affiliation{Max-Planck-Institut f\"ur Kernphysik, 69117 Heidelberg, Germany}}
\newcommand{\munster}{\affiliation{Institut f\"ur Kernphysik, University of M\"unster, 48149 M\"unster, Germany}}
\newcommand{\nikhef}{\affiliation{Nikhef and the University of Amsterdam, Science Park, 1098XG Amsterdam, Netherlands}}
\newcommand{\nyuad}{\affiliation{New York University Abu Dhabi - Center for Astro, Particle and Planetary Physics, Abu Dhabi, United Arab Emirates}}
\newcommand{\purdue}{\affiliation{Department of Physics and Astronomy, Purdue University, West Lafayette, IN 47907, USA}}
\newcommand{\rice}{\affiliation{Department of Physics and Astronomy, Rice University, Houston, TX 77005, USA}}
\newcommand{\stockholm}{\affiliation{Oskar Klein Centre, Department of Physics, Stockholm University, AlbaNova, Stockholm SE-10691, Sweden}}
\newcommand{\subatech}{\affiliation{SUBATECH, IMT Atlantique, CNRS/IN2P3, Nantes Universit\'e, Nantes 44307, France}}
\newcommand{\torino}{\affiliation{INAF-Astrophysical Observatory of Torino, Department of Physics, University  of  Torino and  INFN-Torino,  10125  Torino,  Italy}}
\newcommand{\ucsd}{\affiliation{Department of Physics, University of California San Diego, La Jolla, CA 92093, USA}}
\newcommand{\wis}{\affiliation{Department of Particle Physics and Astrophysics, Weizmann Institute of Science, Rehovot 7610001, Israel}}
\newcommand{\zurich}{\affiliation{Physik-Institut, University of Z\"urich, 8057  Z\"urich, Switzerland}}
\newcommand{\paris}{\affiliation{LPNHE, Sorbonne Universit\'{e}, CNRS/IN2P3, 75005 Paris, France}}
\newcommand{\freiburg}{\affiliation{Physikalisches Institut, Universit\"at Freiburg, 79104 Freiburg, Germany}}
\newcommand{\napels}{\affiliation{Department of Physics ``Ettore Pancini'', University of Napoli and INFN-Napoli, 80126 Napoli, Italy}}
\newcommand{\nagoya}{\affiliation{Kobayashi-Maskawa Institute for the Origin of Particles and the Universe, and Institute for Space-Earth Environmental Research, Nagoya University, Furo-cho, Chikusa-ku, Nagoya, Aichi 464-8602, Japan}}
\newcommand{\laquila}{\affiliation{Department of Physics and Chemistry, University of L'Aquila, 67100 L'Aquila, Italy}}
\newcommand{\tokyo}{\affiliation{Kamioka Observatory, Institute for Cosmic Ray Research, and Kavli Institute for the Physics and Mathematics of the Universe (WPI), University of Tokyo, Higashi-Mozumi, Kamioka, Hida, Gifu 506-1205, Japan}}
\newcommand{\kobe}{\affiliation{Department of Physics, Kobe University, Kobe, Hyogo 657-8501, Japan}}
\newcommand{\kit}{\affiliation{Institute for Astroparticle Physics, Karlsruhe Institute of Technology, 76021 Karlsruhe, Germany}}
\newcommand{\tsinghua}{\affiliation{Department of Physics \& Center for High Energy Physics, Tsinghua University, Beijing 100084, P.R. China}}
\newcommand{\ferrara}{\affiliation{INFN-Ferrara and Dip. di Fisica e Scienze della Terra, Universit\`a di Ferrara, 44122 Ferrara, Italy}}
\newcommand{\groningen}{\affiliation{Nikhef and the University of Groningen, Van Swinderen Institute, 9747AG Groningen, Netherlands}}
\newcommand{\westlake}{\affiliation{Department of Physics, School of Science, Westlake University, Hangzhou 310030, P.R. China}}
\newcommand{\shenzhen}{\affiliation{School of Science and Engineering, The Chinese University of Hong Kong, Shenzhen, Guangdong, 518172, P.R. China}}
\newcommand{\coimbrapoli}{\affiliation{Coimbra Polytechnic - ISEC, 3030-199 Coimbra, Portugal}}
\newcommand{\uniheidelberg}{\affiliation{Physikalisches Institut, Universit\"at Heidelberg, Heidelberg, Germany}}
\newcommand{\roma}{\affiliation{INFN-Roma Tre, 00146 Roma, Italy}}
\newcommand{\bucknell}{\affiliation{Department of Physics \& Astronomy, Bucknell University, Lewisburg, PA, USA}}



\author{E.~Aprile\,\orcidlink{0000-0001-6595-7098}}\columbia
\author{J.~Aalbers\,\orcidlink{0000-0003-0030-0030}}\groningen
\author{K.~Abe\,\orcidlink{0009-0000-9620-788X}}\tokyo
\author{S.~Ahmed Maouloud\,\orcidlink{0000-0002-0844-4576}}\paris
\author{L.~Althueser\,\orcidlink{0000-0002-5468-4298}}\munster
\author{B.~Andrieu\,\orcidlink{0009-0002-6485-4163}}\paris
\author{E.~Angelino\,\orcidlink{0000-0002-6695-4355}}\torino\lngs
\author{D.~Ant\'on~Martin\,\orcidlink{0000-0001-7725-5552}}\chicago
\author{F.~Arneodo\,\orcidlink{0000-0002-1061-0510}}\nyuad
\author{L.~Baudis\,\orcidlink{0000-0003-4710-1768}}\zurich
% ...
\collaboration{XENON Collaboration}\email[]{xenon@lngs.infn.it}\noaffiliation

%

\date{April 25th, 2025}

\begin{abstract}
\noindent
The XENONnT experiment has achieved an exceptionally low $^\text{222}$Rn activity concentration within its inner \SI{5.9}{tonne} liquid xenon detector of \RadonValueSROne, equivalent to about 430 $^\text{222}$Rn atoms per tonne of xenon. This was achieved by active online radon removal via cryogenic distillation after stringent material selection. The achieved $^\text{222}$Rn activity concentration is five times lower than that in other currently operational multi-tonne liquid xenon detectors engaged in dark matter searches. This breakthrough enables the pursuit of various rare event searches that lie beyond the confines of the standard model of particle physics, with world-leading sensitivity. The ultra-low $^\text{222}$Rn levels have diminished the radon-induced background rate in the detector to a point where it is for the first time comparable to the solar neutrino-induced background, which is poised to become the primary irreducible background in liquid xenon-based detectors.
\end{abstract}


{
\let\clearpage\relax
\maketitle
}

\section{Introduction}
\label{sec:introduction}
\noindent
Deep within underground laboratories, massive detectors with ultra-low energy thresholds stand sentinel in the search for dark matter \cite{XENONnT_instrument,XENONnT_WIMP,XENONnT_lowER,AKERIB2020163047_LZ_instr,LZ_wimp_2023,pandax_wimp_2021,darkside50_wimp}, the enigmatically abundant substance constituting nearly \SI{85}{\percent} of the universe's mass. Although its composition remains unknown, various candidate particles are under investigation \cite{BERTONE2005279_DM}. Among these are weakly interacting massive particles (WIMPs), hypothesized to possess masses ranging ${\cal O} (1)$\,GeV/c$^2$ to ${\cal O} (1)$\,TeV/c$^2$ \cite{Roszkowski_2018_WIMP}. While interactions with ordinary matter are expected to be rare, theoretical models predict that WIMPs could occasionally scatter elastically off atomic nuclei, imparting a characteristic recoil energy on the order of a \mbox{few keV}.\linebreak Detecting these faint signals directly requires both exceptional sensitivity and the ability to differentiate them from background events like radioactive decays or cosmic muons.
\begin{figure}[t]
    \centering
    \includegraphics[width=\columnwidth]{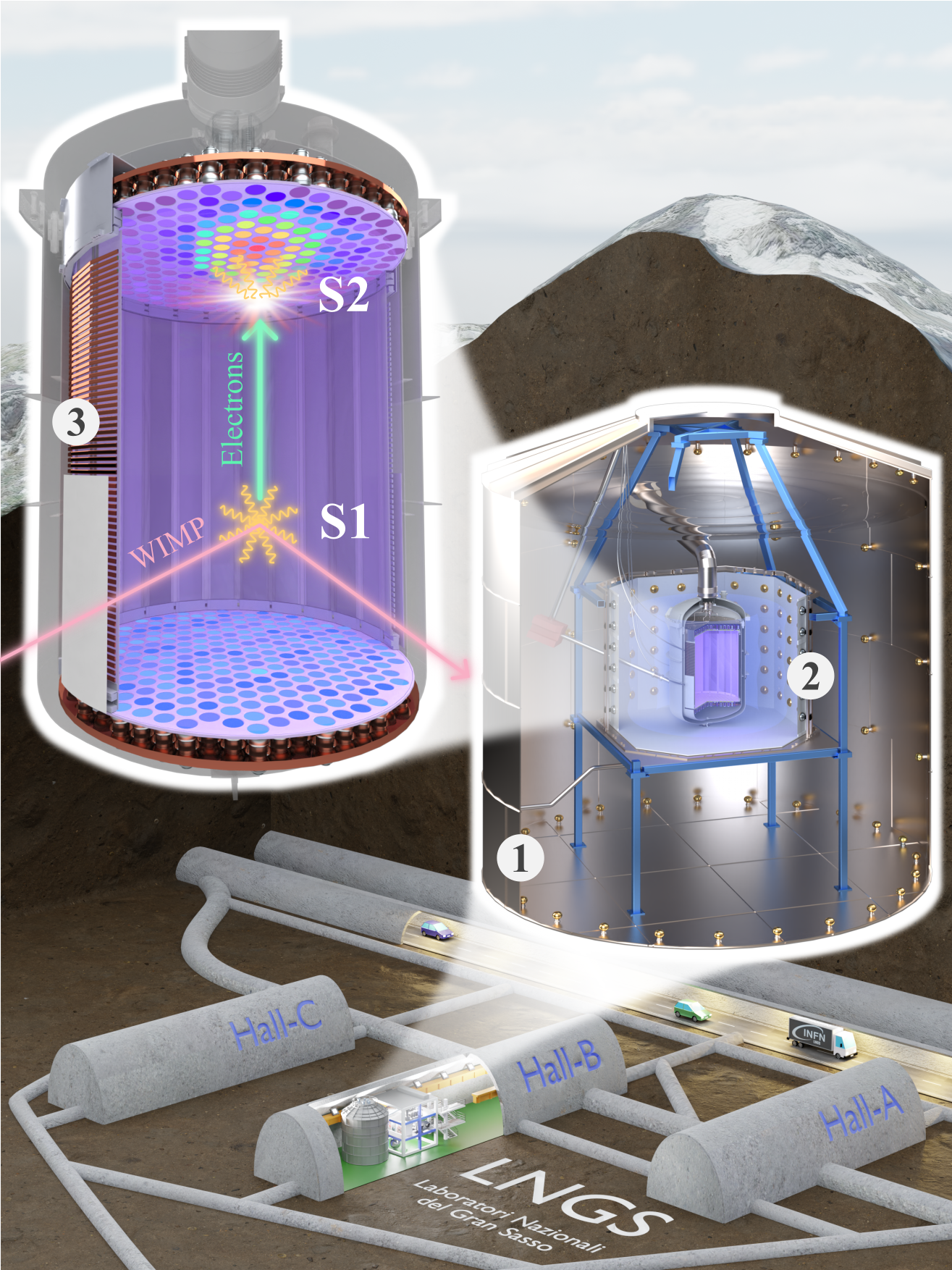}
    \caption{Schematic overview of the XENONnT experiment: The experiment is located \SI{1400}{m} underground at the Laboratori Nazionali del Gran Sasso (LNGS), Italy. This depth significantly reduces cosmic muon and neutron backgrounds due to rock overburden. The experimental setup consists of a service building housing the xenon handling and data acquisition systems, and a water tank containing three nested detectors: (1) The muon veto, (2) the neutron veto, and (3) the Time Projection Chamber (TPC). The TPC, containing \SI{5.9}{t} of liquid xenon, is the central detector, measuring low-energy particle interactions in the LXe volume. Its key features include deposited energy measurement, particle identification, and 3D position reconstruction.}
    \label{fig:tpc}
\end{figure}

Dual-phase xenon time projection chambers (TPCs), pioneered by the XENON collaboration \cite{XENON10,APRILE2012573_xe100_instr,XENON1T_instrument,XENONnT_instrument} and others \cite{ALNER2007287_zeplinII,AKIMOV200746_zeplinIII,AKERIB2013111_LUX_instr,AKERIB2020163047_LZ_instr}, have proven to be highly effective for this purpose \cite{Baudis:2023pzu}. Upon WIMP interaction with the liquid xenon (LXe) target, a prompt scintillation light (S1) is emitted and detected by two photosensor arrays at the top and bottom of the TPC (\autoref{fig:tpc} inset). Additionally, free electrons from the interaction drift upwards in an electric field to the liquid-gas interface. Here, a 20-fold stronger extraction field pulls them into the gaseous xenon (GXe) and accelerates them, inducing electroluminescence: The electrons excite the xenon atoms, producing a delayed and even brighter flash of light (S2), also detected by the photosensors. The location in the horizontal plane is obtained from the S2 signal distribution recorded by the upper photosensor array, while the time difference between S1 and S2, inversely proportional to the known electron drift velocity, provides the vertical coordinate.  By combining the S1 and S2 signals, the deposited energy of an interaction can be precisely reconstructed, allowing for an energy threshold of ${\cal O}(1)$\,keV. Moreover, the S2/S1 ratio helps discriminate WIMP-nucleus scattering (nuclear recoil, NR) from background events like beta/gamma scattering on electrons (electronic recoil, ER). Thanks to the substantial shielding offered by LXe's high density (about three times that of water), radioactivity from detector materials can be largely suppressed by restricting analysis to the central region of the target (fiducialization). This combination of high efficiency, low energy threshold, good energy resolution, and excellent background rejection makes the xenon TPC technology a powerful tool for WIMP and other rare event searches.
    
Its potential, coupled with the captivating nature of the scientific question, has spurred continued advancements. Nearly two decades after the ZEPLIN-II\,\cite{ALNER2007287_zeplinII} and XENON10\,\cite{XENON10} experiments with approximately \SI{10}{kg} xenon targets, three currently operational experiments --- PandaX-4T\,\cite{pandax_wimp_2021}, XENONnT\,\cite{XENONnT_instrument}, and LZ\,\cite{AKERIB2020163047_LZ_instr} --- have emerged. Leveraging active LXe masses ranging from 4 to 7 tonnes, these experiments search for WIMPs and other rare phenomena within underground laboratories across China, Italy, and the USA. Additionally, a similar argon-based experiment, DarkSide-20k \cite{Darkside20k_2018}, is under construction in Italy.

Beyond WIMPs, the xenon TPC technology tackles another equally crucial and timely science question: the search for neutrinoless double beta decay \cite{RevModPhys_95_025002_NDBD}. This elusive process holds the key to unlocking the universe's matter-antimatter asymmetry and the remarkably small mass of neutrinos. Like WIMP searches, it demands exceptional sensitivity and background reduction, but at higher signal energies of a few MeV. The NEXT-100 experiment at the Canfranc Underground Laboratory (Spain) utilizes a high-pressure gaseous xenon TPC to explore this phenomenon\,\cite{NEXT:2015wlq}, employing topological event reconstruction for enhanced background discrimination. This approach offers a complementary strategy to liquid xenon experiments. Building upon the successful EXO-200 experiment\,\cite{M_Auger_2012_exo200_instr}, the nEXO experiment\,\cite{nEXO_Rn222} is planned for the Canadian underground laboratory, SNOLAB. This next-generation detector will contain 5 tonnes of LXe enriched in the isotope \isotope[136]{Xe}, a promising candidate for neutrinoless double beta decay, offering a massive target for the rare decay. As with WIMPs, the decisive factor here is the ultra-low background rate alongside the large isotope mass.

These experiments aim to detect a handful of rare events above the detector background over their operational lifetime. Consequently, minimizing background is crucial and involves multiple steps. Experiments are positioned in underground laboratories to reduce exposure to cosmic muons and are further shielded with active veto systems, typically based on large water or liquid scintillator volumes, to identify muons and neutrons. Stringent material selection ensures minimal radioactivity within the detector itself. After fiducialization and advanced event discrimination techniques, only background sources remain that cannot be shielded or are dissolved in the LXe itself. The first category includes solar and atmospheric neutrinos. The second category includes radioactive isotopes such as $^{3}$H, $^{37}$Ar, $^{39}$Ar, $^{85}$Kr, $^{220}$Rn and \radon dissolved in LXe. Entry points include xenon extraction from air separation, emanation from detector materials, or air leaks. Noble gas impurities cannot be easily mitigated by conventional noble gas purifiers, e.g. high-temperature getters. Additionally, long-lived isotopes like $^{124}$Xe and $^{136}$Xe with half-lives longer than $10^{20}$\,years are present within the xenon. The decay of $^{124}$Xe via double electron capture has been observed for the first time with the XENON1T experiment \cite{XENON1T_ECEC_nature,XENON1T_ECEC0nbb_prc}, XENONnT's predecessor, and offers valuable calibration opportunities for future generations of detectors. The two-neutrino double beta decay of $^{136}$Xe remains a non-negligible background source for dark matter searches \cite{XENONnT_lowER}. 

Previous work by the XENON Collaboration demonstrated effective removal of lighter noble gases like argon and krypton from LXe using cryogenic distillation, achieving negligible concentrations \cite{XENON1T_kr_removal}. However, a one-time removal is not sufficient for radon isotopes, that continuously emanate from detector materials due to the decay chains of primordial uranium and thorium present in virtually all materials.

The longest-lived radon isotope \radon with its half-life of $\RadonHalfLife$ \cite{ENSDF} poses the most significant background challenge for dark matter and neutrinoless double beta decay searches in xenon detectors. Their decay progeny present particular difficulties. For example, the beta decay of the daughter isotope $^{214}$Pb deposits energy as an ER in the TPC. This background cannot be sufficiently eliminated through an S2/S1 ratio cut, which currently achieves \SI{99.3}{\percent} efficiency in XENONnT at a \SI{50}{\percent} NR signal acceptance \cite{XENONnT_WIMP}. Although the closest alpha decay occurs on the order of 30 minutes before or after the beta decay of \isotope[214]{Pb}, there are promising attempts to reconstruct the path of the \isotope[214]{Pb} atom in the LXe using convection and diffusion models, and thus identifying the event as a progeny of \isotope[222]{Rn}~\cite{XENONCollaborationP:2024xwn, LZCollaboration:2024lux}.

Notably, the remaining ER signals cannot be differentiated from potential WIMP interaction signals. Similarly, for the neutrinoless double beta decay search in $^{136}$Xe, the major background arises from the gamma line emitted by another radon progeny, $^{214}$Bi, whose energy falls close to the expected Q-value of the double beta decay.

In XENON1T's main science run, the \radon activity concentration stood at \RadonValueOneTonMain. An R$\&$D run achieved a lower value of \RadonValueOneTonLow through two key innovations: first, the existing GXe purification pumps were exchanged with a novel, nearly radon-free, magnetically-coupled piston pump \cite{Xe1T_magpump}, and second, radon was actively removed by operating the krypton distillation system in inverse mode \cite{XENON1T_radon_emanation}.

The XENONnT experiment demands a \radon activity concentration of \RadonValueDesign, translating to just one \radon atom per 16 moles of xenon, or per \SI{2}{\kilo \gram} of xenon. The XENONnT radon removal strategy is multifaceted: the first line of defense is material selection to minimize radon emanation from the outset, followed by an inherent surface-to-volume advantage of the larger detector volume compared to XENON1T, and finally, active removal to eliminate any remaining radon and to reach the desired level.

Beyond WIMP searches, this exceptionally low \radon level opens doors to a diverse physics program utilizing ER events \cite{XENONnT_lowER}. World-leading sensitivity is attainable for various searches, including solar axions, neutrino magnetic moment, bosonic dark matter (dark photons, ALPs), low-mass WIMPs via the Migdal effect, and low energy ER peak searches. Furthermore, the solar neutrino-induced rate in the detector would match the radon-induced one. Further reducing the \radon activity concentration remains crucial for solar neutrino and double beta decay searches \cite{DARWIN_XLZD_White_Paper} but would provide less improvement for WIMP and low energy ER searches in XENONnT.

This paper focuses on demonstrating the potential of the newly developed cryogenic distillation-based Radon Removal System to continuously keep XENONnT's radon concentration at a sub-\si{\micro \becquerel \per \kilo \gram} level. The removal strategy, the system and the results are detailed in \autoref{sec:radonremoval}. The removal efficiency is directly measurable and quantifiable through \mbox{in-situ} alpha decay measurements using the TPC. This allows for a direct comparison of the various removal modes to the one without removal. The impact on future detectors is discussed in \autoref{sec:impact}, the conclusions are presented in \autoref{sec:conclusions}.
\section{Radon Removal in XENONnT}
\label{sec:radonremoval}
\begin{figure*}[t]
\centering
\includegraphics[width=\textwidth]{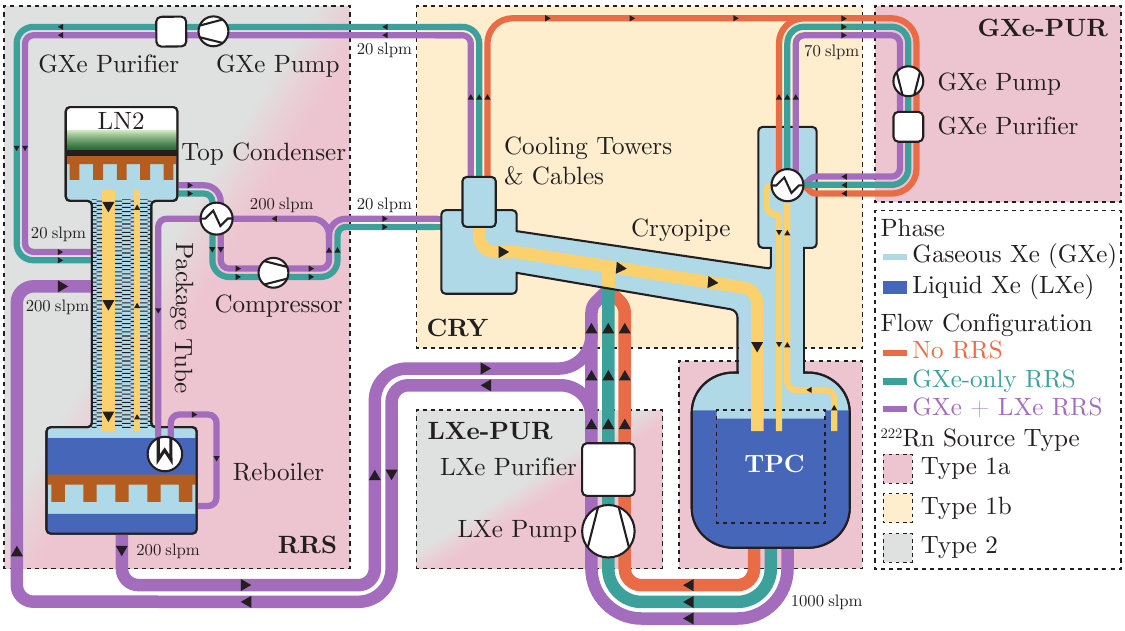}
\caption{Online Radon Removal System and operational modes. The figure depicts the xenon handling systems and their interplay for continuous purification of the xenon inside the Time Projection Chamber (\tpc). Xenon circulation (thin/thick lines for gaseous xenon (GXe)/liquid xenon (LXe)) is maintained via dedicated pumps (GXe/LXe) and purification systems (\gxepur/\lxepur) to remove electronegative impurities. Internal xenon flows are highlighted in yellow. A cryogenic distillation column-based Radon Removal System (\rrs) targets radon emanating from various subsystems categorized by source types 1a (red), 1b (beige), and 2 (gray). Three operational modes are highlighted: In the \noRRS mode (orange), the \rrs is bypassed, and all radon emanated enters the \tpc. In the \GXeOnlyRRS mode (teal), radon-rich GXe is extracted from the cryogenic system (\cry) and directed to the \rrs for purification. The radon-depleted GXe is then returned to the \cry system. In the \GXeLXeRRS mode (violet), additionally a fraction of LXe from the \lxepur is diverted to the \rrs. The resulting radon-depleted GXe is liquefied and fed back into the \lxepur system.}
\label{fig:onlineradonremoval}
\end{figure*}
\noindent
The XENONnT experiment \cite{XENONnT_instrument} is more than just a LXe \tpc housed in a cryostat, surrounded by veto detectors. It is equipped with a variety of xenon handling systems, as shown in \autoref{fig:onlineradonremoval}. A cryogenic system (\cry) is required to maintain thermodynamic equilibrium of the xenon within the detector at a temperature around \SI{-100}{\degreeCelsius}. This \cry system comprises two pulse tube refrigerators and an emergency liquid nitrogen (LN$_2$) cooling tower connected to the inner cryostat via a cryopipe to balance evaporated xenon resulting from external heat input by condensation. Cable feedthrough vessels connected to the cryopipe contain high voltage and signal cables for the photomultiplier tubes (PMTs) inside the detector.

Furthermore, two purification systems are used to continuously clean the xenon. LXe from the detector is extracted and evaporated via a series of heat exchangers before being purified from electronegative impurities using a gas purification system (\gxepur) comprising a radon-free GXe pump and getter-based purifier. In parallel, a novel LXe purification loop (\lxepur) \cite{XENONnT_instrument,Plante:2022khm} circulates LXe through purifiers using a cryogenic liquid pump, also to remove electronegative impurities. The detector's LXe volume containing around \SI{8500}{kg} of xenon can be exchanged once per day by the \lxepur system. While most of the xenon through the \lxepur system is returned to the cryopipe and subsequently to the detector, a fraction is directed to the Radon Removal System (\rrs) to reduce the \radon activity concentration in the detector.

All internal surfaces in contact with GXe or LXe are potential sources of radon, which continuously emanates into the xenon target material. Radon emanation measurements \cite{XENON1T_radon_emanation,radioactivity_XENONnT} reveal the location and magnitude of different radon sources in the XENONnT system. The sources are classified into two main types, depending on their position in the XENONnT system relative to the \rrs. Type 1 sources emanate and are flushed directly into the detector before reaching the \rrs. They are further subdivided into type 1a sources, which go directly into the cryostat's LXe, and type 1b sources, which are within the cryostat's and CRY system's GXe phase. A radon source upstream of the \rrs in the xenon handling system is referred to as type 2. Consequently, type 2 radon enters the \rrs before reaching the LXe inside the detector, and can therefore be efficiently removed. The classification of the different subsystems in XENONnT is highlighted in \autoref{fig:onlineradonremoval}. Subcomponents of a subsystem can contribute to different source types (\autoref{subsec:rn_sources}).

The radon removal system has two modes: In the LXe-mode, a LXe flow is continuously extracted from the detector, purified with the RRS, and fed back as radon-depleted LXe. The high flow of this mode is projected to achieve a two-fold reduction in radon concentration. In the GXe-mode, xenon is extracted from the GXe phase before entering the detector's LXe phase, effectively converting type 1b sources into type 2 sources that go directly to the RRS \cite{XENON1T_radon_emanation}. This mode potentially achieves another factor-of-two reduction in \radon activity concentration depending on the extraction efficiency.
\newpage
\noindent
Three operational modes that were performed in XENONnT are visualized in \autoref{fig:onlineradonremoval}: 
\begin{itemize}
    \item [a)] \textbf{No RRS:} The \rrs system is not in operation (orange).
    \item [b)] \textbf{GXe-only RRS mode}: Only GXe extraction is performed (teal).
        \item [c)] \textbf{GXe+LXe RRS mode:} A combination of GXe and LXe extraction is performed (violet).
\end{itemize}

\subsection{Radon Removal System}
\begin{figure*}[t]
\centering
\includegraphics[width=\textwidth]{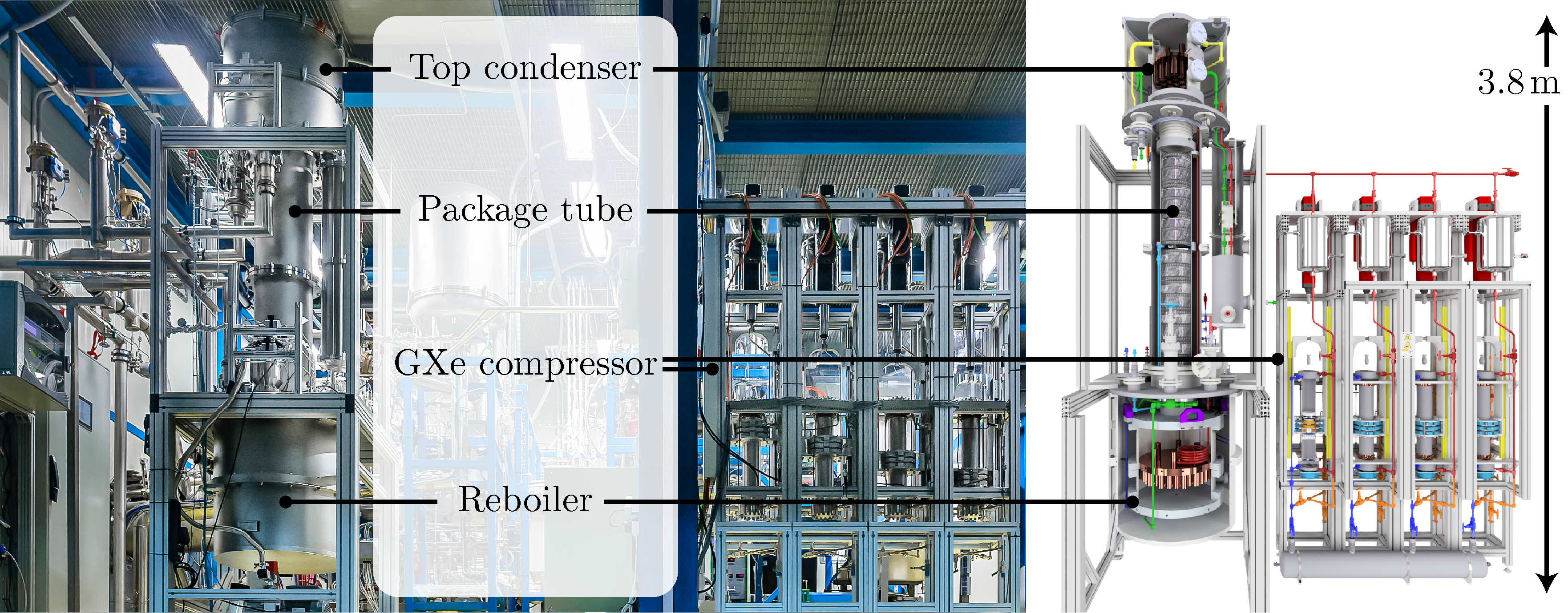}
\caption{Radon removal system installed underground in the service building of the XENONnT experiment at LNGS (left) and related CAD drawing (right). The four key components top condenser, package tube, reboiler and GXe compressor for the heat pump, composed of a four-cylinder magnetically-coupled piston pump, are visible.}
\label{fig:rrs_photo}
\end{figure*}

The XENONnT Radon Removal System, shown in \autoref{fig:rrs_photo}, leverages a cryogenic distillation column \cite{RAD_EPJC}, exploiting the difference in vapor pressure between xenon and radon. Radon reduction is achieved by effectively trapping radon in a LXe reservoir at the bottom of the column. The relatively short half-life of radon (\RadonHalfLife) ensures its decay within the reservoir, eliminating the need for extraction of the impurity-enriched xenon, unlike in conventional distillation systems for krypton or argon removal \cite{XENON1T_kr_removal}.
This feature makes the radon removal process inherently xenon-loss-free allowing for a continuous operation. To significantly reduce radon and mitigate its decay-induced background signals in the LXe TPC, the \rrs is designed to operate mainly with LXe input and output, and the process flow must be sufficiently large to purify the TPC's LXe mass on a timescale comparable to or shorter than the radon mean lifetime (\SI{5.5}{days}). A LXe extraction flow of about \SI{71}{\kilo\gram \per \hour} (\SI{200}{slpm}) through the \rrs would correspond to a two-fold radon reduction for type 1 radon sources given the \SI{8500}{kg} xenon in the cryostat \cite{RAD_EPJC}. 

The distillation tower comprises three key components: a top condenser, a central package tube, and a bottom reboiler. The reboiler houses the LXe reservoir (capacity: up to \SI{130}{kg}) where radon accumulates due to its 10-fold lower vapor pressure compared to xenon at \SI{-100}{\degreeCelsius} \cite{NIST}. The reboiler also employs electrical heaters to vaporize a fraction of the LXe, creating an upward gas stream (flow rate: \SI{106}{\kilo\gram \per \hour} (\SI{300}{slpm})) through the package tube. This tube contains a structured packing material with a large surface area for efficient liquid-gas exchange across its entire height (\SI{190}{cm}), effectively acting as a series of interconnected theoretical distillation stages \cite{RAD_EPJC}.

Radon-rich LXe and GXe from the detector enter the mid-section of the tube. The liquid flows downward to the reboiler, while the gas ascends towards the top condenser. Here, a fraction of the upward gas stream from the reboiler and the feed condenses and flows back to the package tube at a rate of \SI{35}{\kilo\gram \per \hour} (\SI{100}{slpm}), maintaining a LXe reflux ratio of \num{0.5} relative to the \SI{71}{\kilo\gram \per \hour} (\SI{200}{slpm}) radon-depleted xenon extraction flow. This rectification process prevents radon escape from the top.

The system is designed to achieve a 1000-fold radon enrichment between the feed and the reboiler, and a 100-fold reduction between the feed and the top condenser. With the chosen design parameters, including process flow, reflux ratio, and inlet and outlet radon concentrations, the McCabe-Thiele method allowed calculation of the required number of theoretical distillation stages and consequently the total height of the packed column \cite{RAD_EPJC}. Further, the necessary cooling and heating powers were derived to facilitate xenon phase changes throughout the system.

The top condenser requires approximately \SI{1}{kW} of cooling power to maintain the desired reflux ratio of \num{0.5} and is achieved with a custom-made bath-type heat exchanger operated with liquid nitrogen \cite{RAD_HE_JINST}. The extracted radon-depleted GXe flow of \SI{71}{\kilo\gram \per \hour} (\SI{200}{slpm}) from the top condenser must be reliquefied for reinjection into the LXe purification circuit. This additional liquefaction step necessitates another \SI{2}{kW} of cooling power, bringing the total cooling requirement to \SI{3}{kW}. The reboiler, on the other hand, requires \SI{3}{kW} of heating power to generate the upward evaporation flow of \SI{106}{\kilo\gram \per \hour} (\SI{300}{slpm}), vital for a stable distillation process within the package tube.

The substantial cooling power required for the xenon liquefaction at the outlet necessitates an energy-efficient solution. This is achieved through the heat pump principle: gaseous radon-depleted xenon from the top is first compressed with a GXe compressor and is then liquefied in a novel bath-type GXe-LXe heat exchanger (HE) integrated into the reboiler. This HE features two compartments: a top vessel containing LXe and a bottom vessel holding radon-depleted GXe. By thermally connecting these compartments, the GXe condenses in the bottom while LXe evaporates in the top, eliminating the need for the additional \SI{2}{kW} of electric heating. This heat exchange process relies on a temperature difference and thus a pressure difference between the compartments. The GXe pressure must be higher than the LXe pressure, allowing the GXe to condense at a higher temperature and establish the necessary temperature gradient that drives heat transfer. To achieve this pressure differential, a four-cylinder magnetically coupled piston pump is employed as a compressor \cite{RAD_4MP_JINST}. Additional GXe/GXe HEs further optimize the system efficiency by pre-cooling and pre-warming the GXe flow between the tower and the compressor. XENONnT's demanding radiopurity requirements, ensuring the \rrs itself does not contribute significantly to the radon in the detector, necessitated rigorous material screening and custom fabrication of most system components. This included the two HEs at the top and bottom, the package tube, as well as the compressors.

The \rrs system underwent extensive commissioning in an internal bypass configuration, where the \rrs liquid outlet was internally connected with its liquid inlet. Under thermodynamically stable conditions, the system operated at a flow rate of \SI{91(2)}{\kilo\gram \per \hour} (($258\pm6$)\,slpm). This represents a 30\% increase above the design flow rate for the XENONnT operation, validating the system's capacity for extended performance and potential flexibility \cite{RAD_EPJC}.

\subsection{Radon Alpha Decays in XENONnT}
\label{sec:tpcradmonitor}
\begin{figure}[t]
    \centering
    \includegraphics[width=\columnwidth]{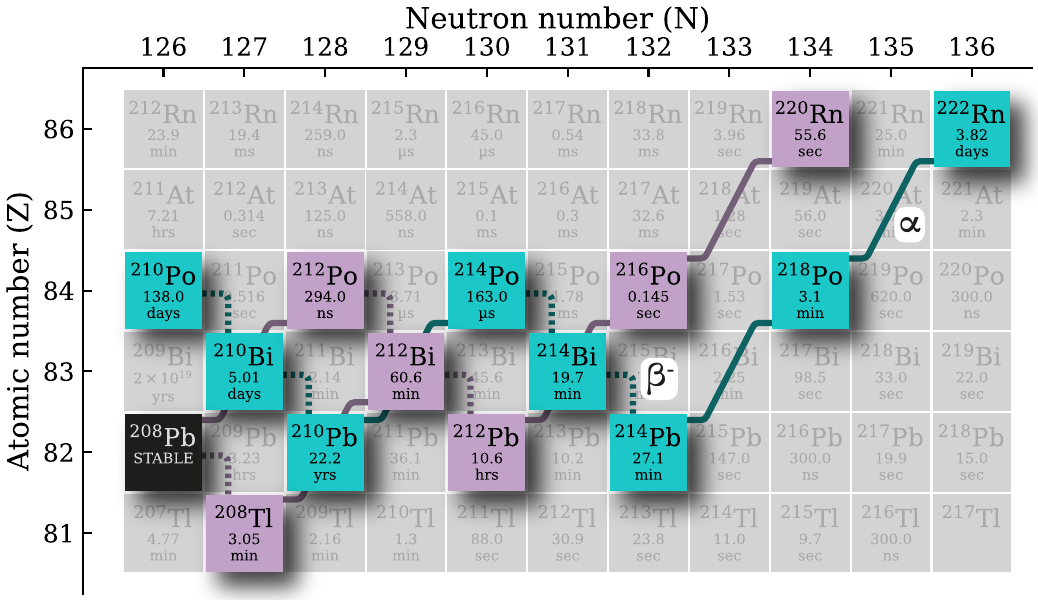}
    \vspace{-0.5\baselineskip}
    \includegraphics[width=\columnwidth]{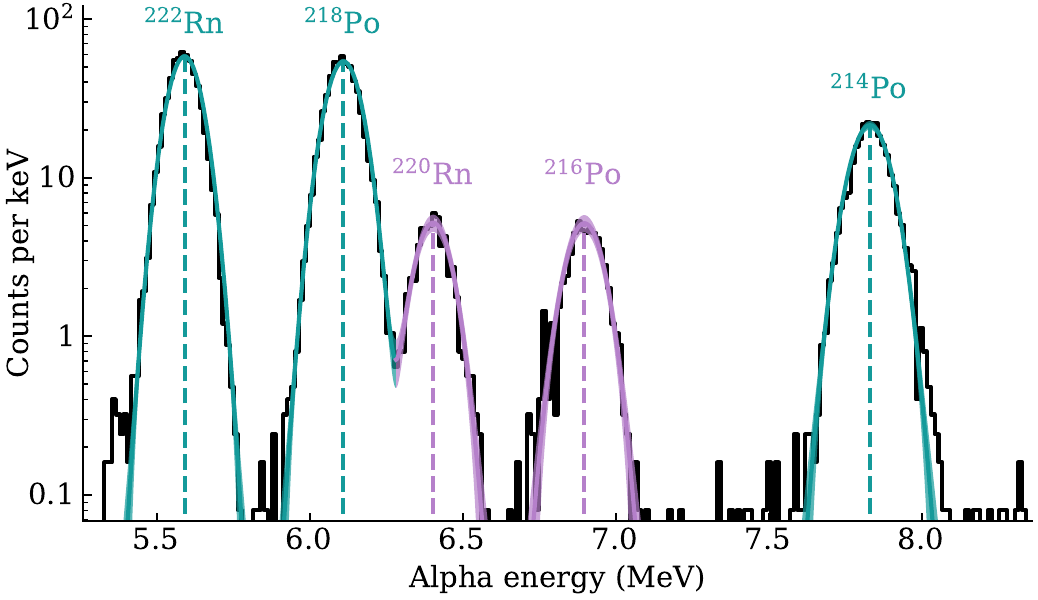}
    \caption{\textbf{Top:} \isotope[222]{Rn} (teal) and \isotope[220]{Rn} (violet) decay chains as part of the primordial uranium and thorium chains. Alpha ($\upalpha$) decays are indicated by solid lines, while beta ($\upbeta^{\text{-}}$) decays are indicated by dashed lines. Note that while the decay chains appear to intersect, each isotope belongs exclusively to either the \isotope[222]{Rn} or the \isotope[220]{Rn} decay chain. All values are taken from Ref. \cite{ENSDF}.
    \textbf{Bottom:} Reconstructed energy spectrum of alpha decays in the XENONnT detector. The relative energy resolution is better than 1\%, allowing to distinguish the alpha decays of different isotopes within the \radon and \isotope[220]{Rn} decay chains. The fit of the data is done with Gaussian functions. Ionized progenies from the decay chains partly plate out on the cathode and detector walls under the influence of the electric field and due to LXe convection. This results in reduced concentrations in the LXe bulk for isotopes further down the decay chain, as shown by the height difference between, for example, \isotope[218]{Po} and \isotope[214]{Po}.}
    \label{fig:spectrum}
\end{figure}
\noindent
Radon is a primordial decay product, arising from both the uranium and thorium decay chains. The isotopes \isotope[219]{Rn} and \radon are produced in the uranium chains of \isotope[235]{U} and \isotope[238]{U}, respectively, while \isotope[220]{Rn} originates from the thorium chain of \isotope[232]{Th}. Due to its short half-life of less than \SI{4}{\second} \cite{ENSDF}, \isotope[219]{Rn} decays before it reaches the central xenon volume and does not pose a background source for the dark matter search. The decay chains starting from the other two isotopes \isotope[220]{Rn} and \isotope[222]{Rn} are highlighted in \autoref{fig:spectrum} (top). The decay chain of \radon includes several alpha and beta emitters. While alpha particles have distinct MeV energies, beta particles possess continuous spectra up to their endpoints of a few hundreds of keV, which can be misinterpreted as low-energy WIMP signals due to the imperfect ER/NR discrimination. Among these emitters, the \radon progeny \isotope[214]{Pb} stands out due to its \SI{9.2}{\percent} branching ratio for beta decay to the ground-state without accompanying gamma emission \cite{TabRad_v4}. This decay mode poses a significant background source since all other beta decays of the uranium and thorium decay chains are either identifiable by time-coincident alpha or gamma decays, or involve decays with longer half-lives, like the one of \isotope[210]{Pb}. While the XENONnT detector is optimized for the detection of low-energy interactions in the keV region, its sensitivity extends well into the MeV range, enabling \mbox{in-situ} measurements of the alpha decays (see \autoref{fig:spectrum}, bottom).
\begin{figure*}[t]
    \centering
    \includegraphics[width=1.0\textwidth]{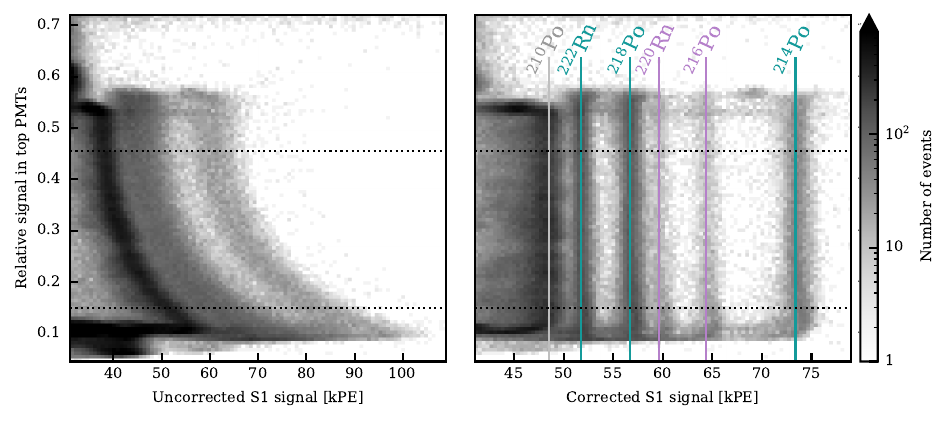} 
    \caption{Scintillation signal of alpha interactions in the XENONnT detector. \textbf{Left:} Depth dependence of the uncorrected light signal. \textbf{Right:} Same data after all geometric corrections from Analysis-I are applied (see text). The horizontal dashed lines indicate the selected subvolume used for the analysis.}\label{fig:s1_aft_dependence}
\end{figure*}

The full decay energy $Q_\upalpha$ of alpha particles is deposited in the LXe. However, due to their high stopping power, alpha particles produce short tracks ($\leq$\SI{100}{\micro\meter})\,\cite{Ziegler:2010}. With around $9\,000$ detected scintillation photons per MeV of deposited energy, alpha decays can be detected using the primary scintillation signal (S1) only. This allows for continuous monitoring of the activity concentration, even when the charge signal (S2) is unavailable, for example, when no extraction field is present.

However, event position reconstruction within the \tpc based solely on S1 exhibits larger systematic uncertainties compared to the standard S1+S2 method\,\cite{XENONnT_analys1}. To address this, two independent \radon activity concentration analyses (I and II) were conducted. Both are outlined in the following, with their main difference being the chosen method for the event position reconstruction.

The S1 light collection efficiency for alpha events varies with their location within the detector due to total internal reflection at the liquid-gas interface and reflections off the detector walls. Spatial dependence corrections were derived using either the monoenergetic alpha decays of \isotope[214]{Po} or \isotope[222]{Rn}. Their signal dependencies along the radial and depth coordinate were fitted using polynomial models. Due to an increased concentration of photoabsorbing impurities in the LXe during the \GXeLXeRRS mode, an additional correction to the observed number of photons per alpha event (less than \SI{5}{\percent}) was applied (see\,\cite{XENON_B8} for further details).

In Analysis-I, the horizontal positions were determined from the center of mass of the light distribution recorded by the top and bottom PMT arrays. The depth was inferred from the fraction of light detected by the top array relative to the total detected light (see \autoref{fig:s1_aft_dependence}), with a lower fraction indicating a deeper event position. Analysis-II employed a convolutional neural network (CNN) algorithm to reconstruct both horizontal and vertical coordinates. It was trained on an independent dataset containing S1+S2-derived event locations\,\cite{Guida:2021elc}.

\autoref{fig:spectrum} (bottom) shows the energy spectrum of alpha particles detected in the LXe using the Analysis-I corrections. As expected, the three alpha emitting isotopes \radon, \isotope[218]{Po}, and \isotope[214]{Po} from the uranium chain are observed.
Additionally, the spectrum shows a subdominant contribution from \isotope[220]{Rn} and \isotope[216]{Po}, as well as from \isotope[212]{Po} (not shown in the figure), which belong to the thorium decay chain.

To mitigate the influence of alpha decays from \isotope[210]{Po}, a progeny of \isotope[210]{Pb} that accumulates on the detector walls and electrodes, both analyses were restricted to an inner volume, containing a LXe mass of \SI{1.22(3)}{\tonne} for Analysis-I and \SI{2.05(6)}{\tonne} for Analysis-II. 
These masses were estimated using a cross-calibration with S1+S2 events, and their uncertainties were obtained as described in Ref.\,\cite{XENONnT_analys1}.
\begin{figure*}[t]
    \centering
    \includegraphics[width=\textwidth]{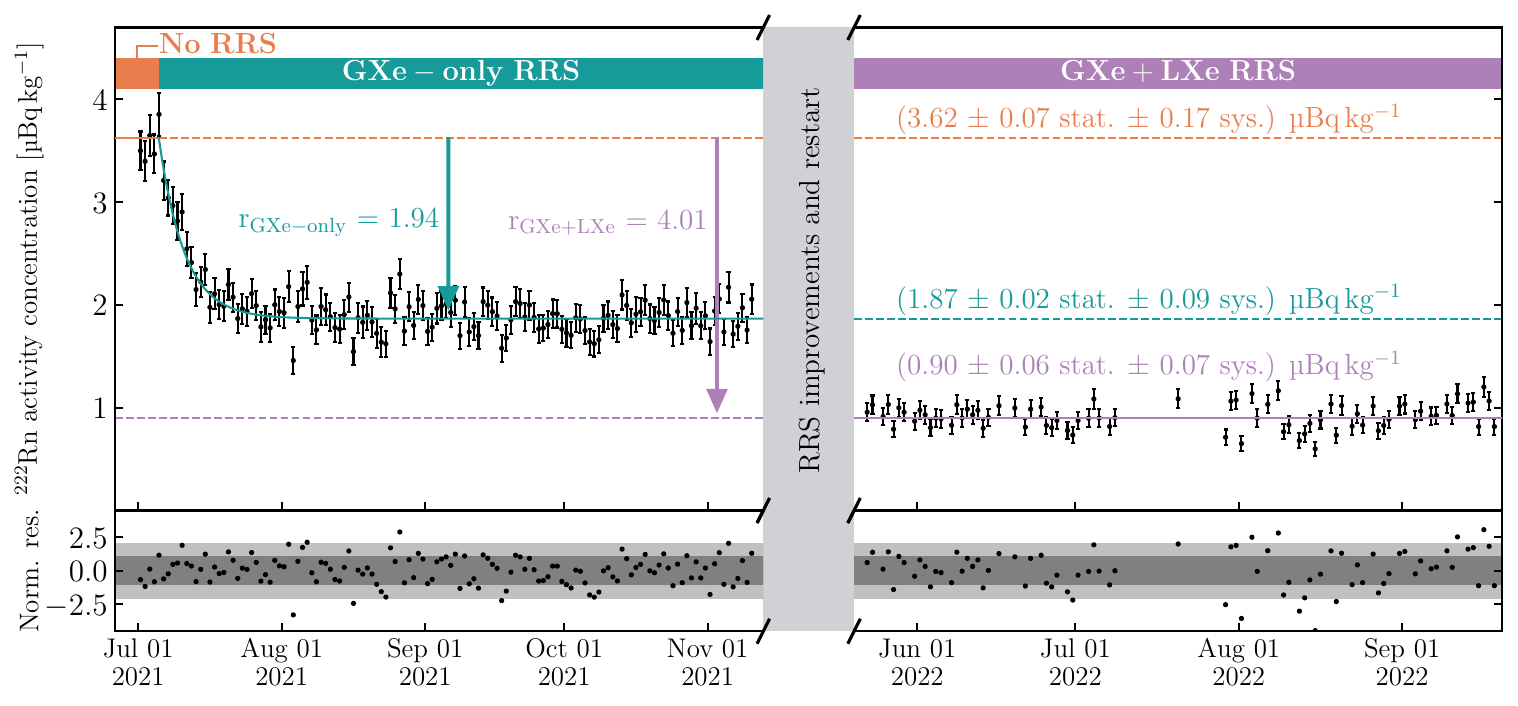}
    \caption{\radon activity concentration (black) in XENONnT as a function of time for the three operational modes: \noRRS (orange), \GXeOnlyRRS (teal), and \GXeLXeRRS (violet). A radon removal model is included as a solid line. Normalized residuals are displayed in the bottom panel. Detector calibrations conducted prior to the presented time period labeled as \noRRS in this work were excluded from the analysis to prevent potential bias in the \radon activity concentration estimate limiting the available data to a few days only. The gray shaded area denotes a period where improvements were made to the Radon Removal System (\rrs) and stable data acquisition was not possible. Radon reduction factors within the Time Projection Chamber (\tpc) are estimated as $r_\mathrm{GXe-only} = \RadonReductionFactorSRZero$ and $r_\mathrm{LXe+GXe} = \RadonReductionFactorSROne$ for \GXeOnlyRRS and \GXeLXeRRS modes, respectively. The minimum \radon activity concentration of \RadonValueSROne is the lowest ever achieved in an operational LXe \tpc. It depends on the \rrs process flow and detector conditions. Data points include statistical uncertainties. The statistical uncertainty in achieved \radon activity concentrations within each mode is calculated from the model fit via bootstrapping, and the systematic uncertainties are derived from the \mbox{in-situ} alpha decay analysis. 
    }
    \label{fig:evolution}
\end{figure*}

Comparing the two analyses, a \SI{1}{\percent} difference was observed during the \noRRS and the \GXeOnlyRRS mode, while a \SI{14}{\percent} difference was noted for the \GXeLXeRRS mode. To account for these differences, the final \radon activity concentration value in each mode was determined as the average of the mean \radon activity concentration derived from both analyses. Additionally, half of the difference between Analysis-I and -II was added to the systematic uncertainty (\SI{0.01}{\micro \becquerel \per \kilo \gram} for the \noRRS and \GXeOnlyRRS as well as \SI{0.06}{\micro \becquerel \per \kilo \gram} for the \GXeLXeRRS mode).

The alpha spectrum shown in \autoref{fig:spectrum} is fitted with a sum of individual Gaussian functions. The fitted peak positions exhibit a linear relationship with the corresponding alpha particle energies and a relative energy resolution of better than \SI{1}{\percent} is achieved. To derive the \radon time evolution shown in \autoref{fig:evolution}, events within a $\pm\,3\,\sigma$ region around the \radon peak were selected. The resulting data was corrected for this selection efficiency and dead-time effects. Approximately one-day time bins were chosen to ensure adequate statistics within each bin. Finally, the measured activity in terms of \si{\micro \becquerel} is divided by the sensitive analysis volume in order to report the activity concentration in terms of \si{\micro \becquerel \per \kg}.
\subsection{Radon Source Distribution and Removal Model}
\label{subsec:rn_sources}

\begin{figure}[t]%
\centering
\includegraphics[width=\columnwidth]{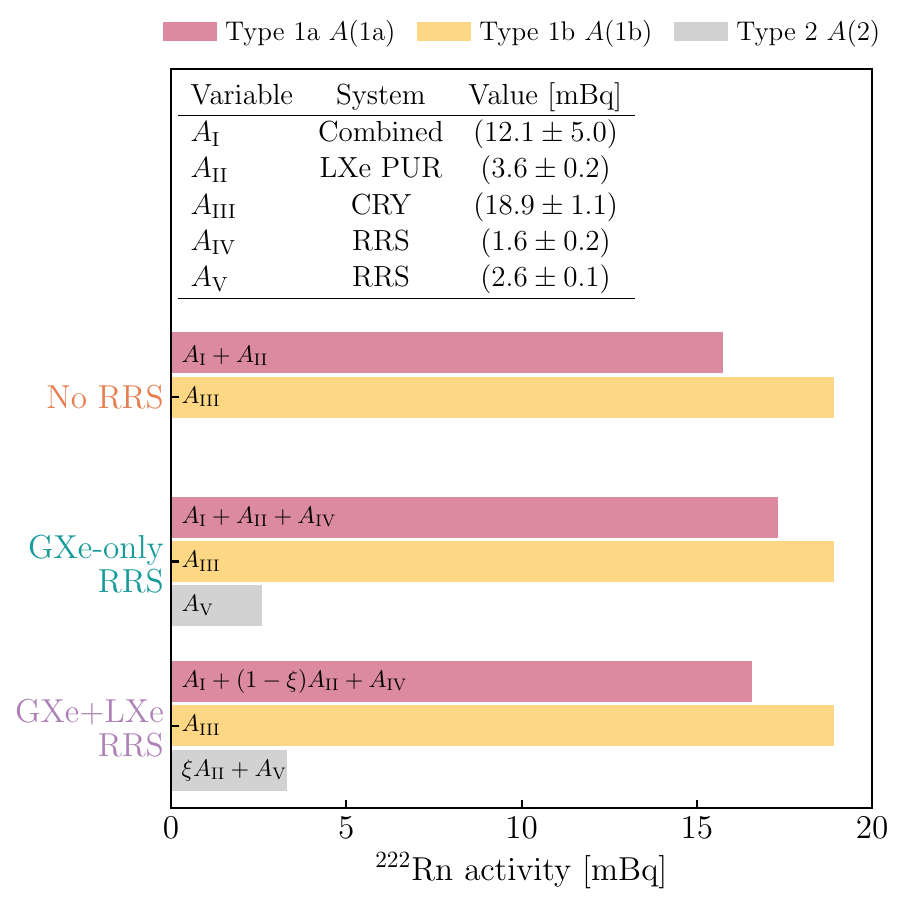}
\caption{Radon source distribution in XENONnT before radon reduction using the \rrs \cite{XENON1T_radon_emanation,radioactivity_XENONnT}. Radon source terms are categorized as type 1a
($A(\mathrm{1a})$, red), type 1b ($A(\mathrm{1b})$, yellow), and type 2 ($A(\mathrm{2})$, gray) varying across the three operational modes. The type 1a sources Cryostat (\num{1.9(2)}), \tpc (\num{8.3(50)}) and \gxepur (\num{1.9(2)}) are combined into term $A_\mathrm{I}$ due to model limitations. The \rrs contributes to both type 1a ($A_\mathrm{IV}$) and type 2 ($A_\mathrm{V}$) sources when in operation. The LXe fraction $\xi$ diverted from the \lxepur to the \rrs during the \GXeLXeRRS mode is set to the design value $\xi = \num{0.2}$. 
}
\label{fig:radonsources}
\end{figure}

\noindent
The contribution of radon sources within XENONnT originating from the different subsystems varies across different radon removal modes (see \autoref{fig:onlineradonremoval}). \autoref{fig:radonsources} provides a comprehensive overview of these sources and their classifications. 

Pure type 1a sources enter directly the LXe in the detector and include the cryostat housing the \tpc, the \tpc itself, and the \gxepur system. These sources are combined into the source term $A_\mathrm{I}$. The \cry system, on the other hand, is a pure type 1b source ($A_\mathrm{III}$), where radon emanates into the GXe above the detector and can be extracted and guided to the RRS with a high efficiency $\epsilon_\mathrm{1b}$ before migrating into the liquid. Type 2 sources are located upstream of the \rrs, and can be almost completely removed before they reach the detector. The classification of the \lxepur system ($A_\mathrm{II}$) depends on the radon removal mode: In the \noRRS and \GXeOnlyRRS modes, it acts as a pure type 1a source. In the \GXeLXeRRS mode, a fraction $\xi \approx 0.2$ of the total LXe flow is diverted to the \rrs and becomes a type 2 source. The remaining fraction $(1-\xi)$ is directly entering the detector and is of type 1a. The \rrs system itself contributes radon only when actively in use. Across both modes, components upstream of the distillation column (e.g., the GXe Pump and GXe Purifier) are categorized as type 2 ($A_\mathrm{V}$) since the emanated radon enters directly the distillation tower, while components downstream (e.g., Compressor) are type 1a ($A_\mathrm{IV}$), as the radon is emanated into the xenon flow after the radon removal process in the distillation tower. 

The \autoref{fig:radonsources} highlights the dominance of type 1 sources in XENONnT, with type 2 sources being minimal. Type 1a and type 1b sources each constitute approximately \SI{50}{\percent} of the total radon contribution. The final \radon activity concentration within the \tpc is influenced by the \rrs's efficiency in reducing these initial sources.

A piecewise continuous function, partitioned according to the three operational modes, was employed to model the data in \autoref{fig:evolution}. During the \noRRS period, the \radon emanation and decay were in equilibrium, resulting in a constant activity concentration, $a_\mathrm{No-RRS}^\mathrm{eq}$, given by 

\begin{equation}
a_\mathrm{No-RRS}^\mathrm{eq} = 
\frac{A(\mathrm{1a}) + A(\mathrm{1b})}{m_\mathrm{Xe}},
    \label{eq:no_rrs_mode}
\end{equation}
where $m_\mathrm{Xe}$ is the xenon mass in the cryostat, and $A(\mathrm{1a}), A(\mathrm{1b})$ represent type 1a and 1b sources (as defined in \autoref{fig:radonsources}). Type 2 sources do not contribute during this period ($A(\mathrm{2}) = 0$).  

A time-dependent function describing the \radon evolution within the detector's LXe volume during the \GXeOnlyRRS or the full \GXeLXeRRS mode was developed in Ref. \cite{RAD_EPJC}. This function considers both radon inflow (via emanation or remnant \radon that was not fully removed by the \rrs) and outflow (via extraction and removal, or decay). 

Following Ref. \cite{RAD_EPJC}, the activity concentration, $a_\mathrm{RRS}(t)$, for this mode is given by
\begin{equation}\label{eq:DGL_solution}
    a_\mathrm{RRS}(t) 
    = \frac{K}{m_{\mathrm{Xe}}} 
    \frac{\lambda_{\mathrm{Rn}}}{\Lambda}  
    + \left(a_\mathrm{RRS}(0) - \frac{K}{m_{\mathrm{Xe}}} \frac{\lambda_{\mathrm{Rn}}}{\Lambda}\right) \cdot e^{-\Lambda \cdot t}
    ,
\end{equation}
with
\begin{align}
K &= A(\mathrm{1a}) + (1 - \epsilon_\mathrm{1b}) A(\mathrm{1b}) + \frac{A(\mathrm{2}) + \epsilon_\mathrm{1b} A(\mathrm{1b})}{R_\mathrm{RRS}}, \\
    \Lambda & = \left( \lambda_{\mathrm{Rn}} + \frac{F_{\mathrm{RRS}}}{m_{\mathrm{Xe}}} \left(1-\frac{1}{R_\mathrm{RRS}}\right)\right),
\end{align}
where $\lambda_{\mathrm{Rn}} = \SI{0.18}{\per \day}$ is the \radon decay constant, $a_\mathrm{RRS}(0)$ is the initial activity concentration at the start of the mode, and $\epsilon_\mathrm{1b}$ is the efficiency of extracting type 1b sources from the CRY system. The ratio of the \rrs process flow $F_\mathrm{RRS}$ and the xenon mass $m_{\mathrm{Xe}}$ characterizes the \rrs purification timescale. The \rrs reduction factor $R_\mathrm{RRS}$, defined as the inlet-to-outlet radon concentration ratio, is assumed constant and independent from the radon concentration.
    
For long enough times ($t\cdot \Lambda\gg 1$), equilibrium is reached and the second term of \autoref{eq:DGL_solution} vanishes, leading to $a_\mathrm{RRS}(\infty) = K/m_{\mathrm{Xe}} \cdot \lambda_{\mathrm{Rn}} /\Lambda$ for the \GXeLXeRRS mode.
During the \GXeOnlyRRS mode, LXe extraction is not active ($F_\mathrm{RRS} = 0$). The activity concentration is then described by \autoref{eq:DGL_solution} with $\Lambda = \lambda_{\mathrm{Rn}}$, and the equilibrium plateaus at $a_\mathrm{RRS}(\infty) = K /m_{\mathrm{Xe}}$. Please note the different source terms $A(\mathrm{1a}), A(\mathrm{1b}$), and $A(\mathrm{2})$ in $K$ for both modes as defined in \autoref{fig:radonsources}.

A $\chi^2$-fit was performed to determine model parameters with the following constraints: \radon source terms $A_\mathrm{I}$ to $A_\mathrm{V}$ as defined in \autoref{fig:radonsources}, xenon mass $m_{\mathrm{Xe}} = \SI{8520(85)}{kg}$ and \rrs process flow $F_\mathrm{RRS} = \SI{62(6)}{\kilo\gram \per \hour}$, slightly below the design value for this campaign. The \rrs reduction factor $R_\mathrm{RRS}$ and the \radon extraction efficiency $\epsilon_\mathrm{1b}$ from the GXe were free parameters. The best-fit yielded $\epsilon_\mathrm{1b} = (\ExtractionFitVal)$. Given the dominant role of the process flow $F_\mathrm{RRS}$ in the removal efficiency \cite{RAD_EPJC}, the model exhibits limited sensitivity to large $R_\mathrm{RRS}$ values, necessitating a one-sided confidence interval $ R_\mathrm{RRS}> \RRRSFitVal$ (\SI{90}{\percent} C.L.), clearly exceeding the design value of $R_\mathrm{RRS} \geq 100$. 

\radon reduction factors in the detector for the \GXeOnlyRRS ($r_\mathrm{GXe\text{-}only}$) and \GXeLXeRRS ($r_\mathrm{GXe+LXe}$) modes as shown in \autoref{fig:evolution} were determined by comparing the plateau activity concentrations to the initial \noRRS plateau.

\autoref{fig:evolution} depicts the \radon activity concentration over time, encompassing the three periods corresponding to the \noRRS (orange), \GXeOnlyRRS (teal), and \GXeLXeRRS (violet) modes. Without the implementation of any active \rrs, the activity concentration is \RadonValuePreSRZero, which is similar to the estimated value of \RadonValueExpectation, inferred from emanation measurements carried out at room temperature \cite{XENON1T_radon_emanation}. During the initial science run of XENONnT \cite{XENONnT_lowER,XENONnT_WIMP}, the \GXeOnlyRRS mode yielded an activity concentration of \RadonValueSRZero. For the second science run, the full extent of the radon removal capabilities was employed, resulting in a remarkable reduction of the activity concentration to \RadonValueSROne, better than the desired design value of \RadonValueDesign. This represents a substantial decrease of a factor of \RadonReductionFactorSROne compared to the initial concentration observed under the \noRRS mode. 
\section{Impact for Future Detectors}
\label{sec:impact}

\begin{figure}[t]%
\centering
\includegraphics[width=\columnwidth]{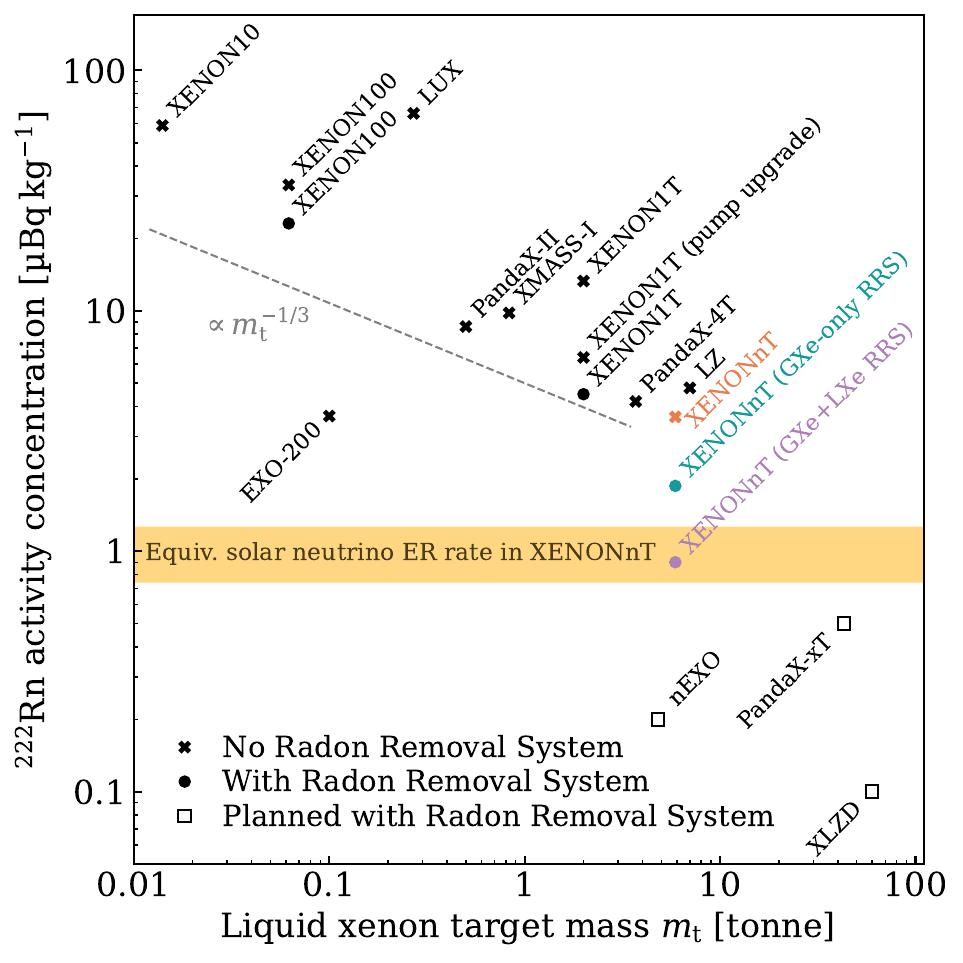}
\caption{\radon activity concentration in LXe-based experiments as a function of their LXe target mass. Crosses indicate experiments without a dedicated Radon Removal System (\rrs), while full circles represent those with active \rrs. Open squares depict projected/required values for future experiments. The gray dashed line visualizes a surface-to-volume improvement trend scaling with $m_\mathrm{t}^{-1/3}$: As detector size increases, the target volume scales cubically, while the radon-emitting surface scales quadratically, leading to a lower \radon activity concentration within the target volume. Experiments with active \rrs, like XENONnT (teal, violet) demonstrably deviate from the overall trend, achieving significantly lower levels. EXO-200, searching for neutrinoless double beta decay, is an exception, achieving a significantly lower \radon level compared to similar-sized experiments by the extremely careful material selection, but probably also due to its liquid-only phase detector design, which minimized type 1 \radon sources directly impacting the active target mass. The yellow band represents the equivalent solar neutrino-induced ER rate in XENONnT (in units of \si{\micro \becquerel \per \kilo \gram}), considering uncertainties in current solar neutrino-xenon interaction models. Data sources include XENON10 \cite{XENON10}, XENON100 \cite{Aprile:2017radon_xe100}, EXO-200 \cite{EXO200_Rn222}, LUX \cite{LUX_Rn222}, PandaX-II \cite{PandaX_II_Rn222}, XMASS-I \cite{XMASS_Rn222}, XENON1T \cite{XENON1T_radon_emanation}, PandaX-4T \cite{pandax_wimp_2021}, XENONnT (this work), LZ \cite{LZ_Rn222}, nEXO \cite{nEXO_Rn222}, PandaX-xT \cite{PandaX:2024oxq}, and XLZD \cite{XLZD:2024gxx}.}
\label{fig:radonhistory}
\end{figure}

\noindent
LXe-based dark matter experiments have witnessed a significant increase in target mass from approximately \SI{10}{kg} two decades ago to the current range of several tonnes. Next-generation detectors are under design to reach up to 60 tonnes or more target mass \cite{DARWIN_XLZD_White_Paper,XLZD:2024gxx}. \autoref{fig:radonhistory} presents the evolution of \radon activity concentration across various experiments,  including those searching for neutrinoless double beta decay, as a function of target mass. As detector target mass steadily increased over time, the \radon activity concentration exhibited a decrease only slightly steeper than expected from the surface-to-volume ratio improvement in larger detectors. Nevertheless, a substantial improvement was only achieved when active removal was implemented, as demonstrated in XENONnT.

In comparison to other tonne-scale detectors currently in operation, a \RadonReductionWrtLZ-fold reduction was accomplished in XENONnT. In comparison to the \radon activity concentration in XENON1T of \RadonValueOneTonMain during its main science runs, an even \RadonReductionWrtOneTon-fold reduction was achieved.  Note that the ER background due to the \isotope[214]{Pb} beta decay can be further mitigated through the implementation of a radon tagging analysis~\cite{XENONCollaborationP:2024xwn, LZCollaboration:2024lux}.

In order to demonstrate the impact of the radon reduction technique on solar neutrino measurements, the expected solar neutrino rate was converted into an equivalent \radon activity concentration. This denotes the \radon concentration, at which the rate of \isotope[214]{Pb} beta decays matches the expected rate of solar neutrino-induced electronic recoils within the energy interval of [5, 30] keV. The rates were computed using the calculation in Ref.\,\cite{CHEN2017656_RRPA}, which assumed that 55\,\% of the solar pp neutrinos interact as electron neutrinos, and the remaining 45\,\% interact as muon or tau neutrinos, only through neutral-current, as a consequence of neutrino oscillation. Two models for the weak elastic neutrino-electron scattering in LXe were used: A free electron approximation including the stepping of atomic shells (FEA) \cite{CHEN2017656_RRPA} and an ab initio many-body method, the relativistic random phase approximation (RRPA) \cite{CHEN2017656_RRPA}. The RRPA model predicts a \SI{23}{\percent} lower rate due to atomic binding effects. 

The corresponding rate of \isotope[214]{Pb} beta decays within [5, 30] keV, required for the conversion from ER event rate to \radon activity concentration, cannot be predicted a priori as it depends on detector specific effects. This is because ionized radon progenies from the \radon decay chain may partly plate out on the cathode and detector walls under the influence of the electric field and due to LXe convection \cite{Aprile:2017radon_xe100}. This results in reduced concentrations in the LXe bulk for isotopes further down the decay chain. This effect can be seen for the XENONnT data in \autoref{fig:spectrum}, where the contribution of \isotope[214]{Po} is reduced to approximately \SI{40}{\percent} of the \radon one. Similarly, the activity concentration of \isotope[214]{Pb} decays is reduced to approximately \SI{70}{\percent}. The latter was determined by a calibration dataset employing a \radon source \cite{MPIK_Rn222_source}. During this calibration the low energy ER spectrum is dominated by the \isotope[214]{Pb} beta decays, rendering all other contributions from \isotope[85]{Kr}, \isotope[136]{Xe} or solar neutrinos negligible. Note that this ratio is specific for XENONnT and different experiments feature different plate-out effects, leading to a different equivalent solar neutrino-induced rate. The resulting equivalent \radon activity concentrations, expressed in \si{\micro \becquerel \per \kilo \gram}, are shown as a yellow band in \autoref{fig:radonhistory}, with the upper and lower limits determined by the FEA and RRPA models, respectively.
\section{Conclusions}
\label{sec:conclusions}
The reported reduction of the \radon activity concentration by XENONnT to \RadonValueSROne is a critical milestone in low-energy rare event experiments: The \radon level achieved is so low that the background induced by the unshieldable solar neutrinos is for the first time comparable to the \radon-induced background. Solar neutrinos, generated in the Sun's core via fusion reactions, produce a particle flux of approximately 6.5$\times 10^{10}\,\si{\per \second \per \square \centi \meter}$ at Earth\,\cite{solar_neutrino_borexino}. The majority originate from the proton-proton chain and exhibit characteristic energy spectra extending to several MeV \cite{solar_neutrino_borexino}. While predominantly traversing LXe detectors without interaction, a subset induce NRs through coherent elastic neutrino-nucleus scattering (\cevns), mimicking WIMP signals. Recent \cevns measurements of solar $^{8}$B neutrinos by XENONnT and PandaX-4T \cite{XENON_B8, pandaX_B8} underscore the importance of this background. Additionally, solar neutrinos can scatter elastically off electrons, creating low-energy ERs. Solar neutrino ER interactions, previously obscured by the dominant \isotope[214]{Pb} background, could now be directly observed in LXe detectors. This fact is illustrated by the yellow band in \autoref{fig:radonhistory}, which denotes an equivalent solar neutrino-induced rate in XENONnT in units of \si{\micro \becquerel \per \kilo \gram}, encompassing the uncertainties associated with current solar neutrino interaction models with xenon.
XENONnT's groundbreaking achievement of an ultra-low radon level below \SI{1}{\micro \becquerel \per \kilo \gram} opens exciting avenues for exploration beyond the Standard Model such as the search for elusive particles like solar axions, axion-like particles, dark photons, and an enhanced neutrino magnetic moment \cite{XENONnT_lowER}. Furthermore, the successful development and implementation of XENONnT's high-flow Radon Removal System paves the way for future detectors like PandaX-xT\,\cite{PandaX:2024oxq} or XLZD~\cite{XLZD:2024gxx} with 40 to 60 tonnes of LXe. 
Such an experiment holds immense potential as the ultimate observatory for low-energy astroparticle physics \cite{DARWIN_XLZD_White_Paper}. Its capabilities will range from precise solar neutrino studies through neutrino-electron and coherent elastic neutrino nucleus scattering, to searches for extremely rare processes like double weak decays of the xenon isotopes \isotope[124]{Xe} and \isotope[136]{Xe}, particularly its neutrinoless double beta decay. Additionally, it could probe a wide range of dark matter candidates, including WIMPs, down to the elusive neutrino fog --- the theoretical limit where the dark matter signal becomes indistinguishable from the irreducible background of solar and atmospheric neutrinos~\cite{neutrino_fog}.
\section*{Acknowledgments}
We gratefully acknowledge support from the National Science Foundation, Swiss National Science Foundation, German Ministry for Education and Research, Max Planck Gesellschaft, Deutsche Forschungsgemeinschaft, Helmholtz Association, Dutch Research Council (NWO), Fundacao para a Ciencia e Tecnologia, Weizmann Institute of Science, Binational Science Foundation, R\'egion des Pays de la Loire, Knut and Alice Wallenberg Foundation, Kavli Foundation, JSPS Kakenhi, JST FOREST Program, and ERAN in Japan, Tsinghua University Initiative Scientific Research Program, DIM-ACAV+ R\'egion Ile-de-France, and Istituto Nazionale di Fisica Nucleare. This project has received funding/support from the European Union's Horizon 2020 research and innovation program under the Marie Sk\l{}odowska-Curie grant agreement No 860881-HIDDeN. We gratefully acknowledge support for providing computing and data-processing resources of the Open Science Pool and the European Grid Initiative, at the following computing centers: the CNRS/IN2P3 (Lyon - France), the Dutch national e-infrastructure with the support of SURF Cooperative, the Nikhef Data-Processing Facility (Amsterdam - Netherlands), the INFN-CNAF (Bologna - Italy), the San Diego Supercomputer Center (San Diego - USA) and the Enrico Fermi Institute (Chicago - USA). We acknowledge the support of the Research Computing Center (RCC) at The University of Chicago for providing computing resources for data analysis.
We thank the INFN Laboratori Nazionali del Gran Sasso for hosting and supporting the XENON project.

\bibliography{sn-bibliography}   

\begin{thebibliography}{54}%
\makeatletter
\providecommand \@ifxundefined [1]{%
 \@ifx{#1\undefined}
}%
\providecommand \@ifnum [1]{%
 \ifnum #1\expandafter \@firstoftwo
 \else \expandafter \@secondoftwo
 \fi
}%
\providecommand \@ifx [1]{%
 \ifx #1\expandafter \@firstoftwo
 \else \expandafter \@secondoftwo
 \fi
}%
\providecommand \natexlab [1]{#1}%
\providecommand \enquote  [1]{``#1''}%
\providecommand \bibnamefont  [1]{#1}%
\providecommand \bibfnamefont [1]{#1}%
\providecommand \citenamefont [1]{#1}%
\providecommand \href@noop [0]{\@secondoftwo}%
\providecommand \href [0]{\begingroup \@sanitize@url \@href}%
\providecommand \@href[1]{\@@startlink{#1}\@@href}%
\providecommand \@@href[1]{\endgroup#1\@@endlink}%
\providecommand \@sanitize@url [0]{\catcode `\\12\catcode `\$12\catcode
  `\&12\catcode `\#12\catcode `\^12\catcode `\_12\catcode `\%12\relax}%
\providecommand \@@startlink[1]{}%
\providecommand \@@endlink[0]{}%
\providecommand \url  [0]{\begingroup\@sanitize@url \@url }%
\providecommand \@url [1]{\endgroup\@href {#1}{\urlprefix }}%
\providecommand \urlprefix  [0]{URL }%
\providecommand \Eprint [0]{\href }%
\providecommand \doibase [0]{http://dx.doi.org/}%
\providecommand \selectlanguage [0]{\@gobble}%
\providecommand \bibinfo  [0]{\@secondoftwo}%
\providecommand \bibfield  [0]{\@secondoftwo}%
\providecommand \translation [1]{[#1]}%
\providecommand \BibitemOpen [0]{}%
\providecommand \bibitemStop [0]{}%
\providecommand \bibitemNoStop [0]{.\EOS\space}%
\providecommand \EOS [0]{\spacefactor3000\relax}%
\providecommand \BibitemShut  [1]{\csname bibitem#1\endcsname}%
\let\auto@bib@innerbib\@empty
\bibitem [{\citenamefont {Aprile}\ \emph
  {et~al.}(2024{\natexlab{a}})\citenamefont {Aprile} \emph
  {et~al.}}]{XENONnT_instrument}%
  \BibitemOpen
  \bibfield  {author} {\bibinfo {author} {\bibfnamefont {E.}~\bibnamefont
  {Aprile}} \emph {et~al.} (\bibinfo {collaboration} {XENON}),\ }\bibfield
  {title} {\enquote {\bibinfo {title} {{The XENONnT dark matter experiment}},}\
  }\href {\doibase 10.1140/epjc/s10052-024-12982-5} {\bibfield  {journal}
  {\bibinfo  {journal} {Eur. Phys. J.}\ }\textbf {\bibinfo {volume} {C84}},\
  \bibinfo {pages} {784} (\bibinfo {year} {2024}{\natexlab{a}})},\ \Eprint
  {http://arxiv.org/abs/2402.10446} {arXiv:2402.10446 [physics.ins-det]}
  \BibitemShut {NoStop}%
\bibitem [{\citenamefont {Aprile}\ \emph {et~al.}(2023)\citenamefont {Aprile}
  \emph {et~al.}}]{XENONnT_WIMP}%
  \BibitemOpen
  \bibfield  {author} {\bibinfo {author} {\bibfnamefont {E.}~\bibnamefont
  {Aprile}} \emph {et~al.} (\bibinfo {collaboration} {XENON}),\ }\bibfield
  {title} {\enquote {\bibinfo {title} {{First Dark Matter Search with Nuclear
  Recoils from the XENONnT Experiment}},}\ }\href {\doibase
  10.1103/PhysRevLett.131.041003} {\bibfield  {journal} {\bibinfo  {journal}
  {Phys. Rev. Lett.}\ }\textbf {\bibinfo {volume} {131}},\ \bibinfo {pages}
  {041003} (\bibinfo {year} {2023})},\ \Eprint
  {http://arxiv.org/abs/2303.14729} {arXiv:2303.14729 [hep-ex]} \BibitemShut
  {NoStop}%
\bibitem [{\citenamefont {Aprile}\ \emph
  {et~al.}(2022{\natexlab{a}})\citenamefont {Aprile} \emph
  {et~al.}}]{XENONnT_lowER}%
  \BibitemOpen
  \bibfield  {author} {\bibinfo {author} {\bibfnamefont {E.}~\bibnamefont
  {Aprile}} \emph {et~al.} (\bibinfo {collaboration} {XENON}),\ }\bibfield
  {title} {\enquote {\bibinfo {title} {{Search for New Physics in Electronic
  Recoil Data from XENONnT}},}\ }\href {\doibase
  10.1103/PhysRevLett.129.161805} {\bibfield  {journal} {\bibinfo  {journal}
  {Phys. Rev. Lett.}\ }\textbf {\bibinfo {volume} {129}},\ \bibinfo {pages}
  {161805} (\bibinfo {year} {2022}{\natexlab{a}})},\ \Eprint
  {http://arxiv.org/abs/2207.11330} {arXiv:2207.11330 [hep-ex]} \BibitemShut
  {NoStop}%
\bibitem [{\citenamefont {Akerib}\ \emph {et~al.}(2020)\citenamefont {Akerib}
  \emph {et~al.}}]{AKERIB2020163047_LZ_instr}%
  \BibitemOpen
  \bibfield  {author} {\bibinfo {author} {\bibfnamefont {D.S.}\ \bibnamefont
  {Akerib}} \emph {et~al.},\ }\bibfield  {title} {\enquote {\bibinfo {title}
  {{The LUX-ZEPLIN (LZ) experiment}},}\ }\href {\doibase
  10.1016/j.nima.2019.163047} {\bibfield  {journal} {\bibinfo  {journal}
  {Nuclear Instruments and Methods in Physics Research Section A: Accelerators,
  Spectrometers, Detectors and Associated Equipment}\ }\textbf {\bibinfo
  {volume} {953}},\ \bibinfo {pages} {163047} (\bibinfo {year}
  {2020})}\BibitemShut {NoStop}%
\bibitem [{\citenamefont {Aalbers}\ \emph
  {et~al.}(2023{\natexlab{a}})\citenamefont {Aalbers} \emph
  {et~al.}}]{LZ_wimp_2023}%
  \BibitemOpen
  \bibfield  {author} {\bibinfo {author} {\bibfnamefont {J.}~\bibnamefont
  {Aalbers}} \emph {et~al.},\ }\bibfield  {title} {\enquote {\bibinfo {title}
  {{First Dark Matter Search Results from the LUX-ZEPLIN (LZ) Experiment}},}\
  }\href {\doibase 10.1103/PhysRevLett.131.041002} {\bibfield  {journal}
  {\bibinfo  {journal} {Phys. Rev. Lett.}\ }\textbf {\bibinfo {volume} {131}},\
  \bibinfo {pages} {041002} (\bibinfo {year} {2023}{\natexlab{a}})}\BibitemShut
  {NoStop}%
\bibitem [{\citenamefont {Meng}\ \emph {et~al.}(2021)\citenamefont {Meng} \emph
  {et~al.}}]{pandax_wimp_2021}%
  \BibitemOpen
  \bibfield  {author} {\bibinfo {author} {\bibfnamefont {Yue}\ \bibnamefont
  {Meng}} \emph {et~al.},\ }\bibfield  {title} {\enquote {\bibinfo {title}
  {{Dark Matter Search Results from the PandaX-4T Commissioning Run}},}\ }\href
  {\doibase 10.1103/PhysRevLett.127.261802} {\bibfield  {journal} {\bibinfo
  {journal} {Phys. Rev. Lett.}\ }\textbf {\bibinfo {volume} {127}},\ \bibinfo
  {pages} {261802} (\bibinfo {year} {2021})}\BibitemShut {NoStop}%
\bibitem [{\citenamefont {Agnes}\ \emph {et~al.}(2018)\citenamefont {Agnes}
  \emph {et~al.}}]{darkside50_wimp}%
  \BibitemOpen
  \bibfield  {author} {\bibinfo {author} {\bibfnamefont {P.}~\bibnamefont
  {Agnes}} \emph {et~al.},\ }\bibfield  {title} {\enquote {\bibinfo {title}
  {{DarkSide-50 532-day dark matter search with low-radioactivity argon}},}\
  }\href {\doibase 10.1103/PhysRevD.98.102006} {\bibfield  {journal} {\bibinfo
  {journal} {Phys. Rev. D}\ }\textbf {\bibinfo {volume} {98}},\ \bibinfo
  {pages} {102006} (\bibinfo {year} {2018})}\BibitemShut {NoStop}%
\bibitem [{\citenamefont {Bertone}\ \emph {et~al.}(2005)\citenamefont
  {Bertone}, \citenamefont {Hooper},\ and\ \citenamefont
  {Silk}}]{BERTONE2005279_DM}%
  \BibitemOpen
  \bibfield  {author} {\bibinfo {author} {\bibfnamefont {Gianfranco}\
  \bibnamefont {Bertone}}, \bibinfo {author} {\bibfnamefont {Dan}\ \bibnamefont
  {Hooper}}, \ and\ \bibinfo {author} {\bibfnamefont {Joseph}\ \bibnamefont
  {Silk}},\ }\bibfield  {title} {\enquote {\bibinfo {title} {{Particle dark
  matter: evidence, candidates and constraints}},}\ }\href {\doibase
  10.1016/j.physrep.2004.08.031} {\bibfield  {journal} {\bibinfo  {journal}
  {Physics Reports}\ }\textbf {\bibinfo {volume} {405}},\ \bibinfo {pages}
  {279--390} (\bibinfo {year} {2005})}\BibitemShut {NoStop}%
\bibitem [{\citenamefont {Roszkowski}\ \emph {et~al.}(2018)\citenamefont
  {Roszkowski}, \citenamefont {Sessolo},\ and\ \citenamefont
  {Trojanowski}}]{Roszkowski_2018_WIMP}%
  \BibitemOpen
  \bibfield  {author} {\bibinfo {author} {\bibfnamefont {Leszek}\ \bibnamefont
  {Roszkowski}}, \bibinfo {author} {\bibfnamefont {Enrico~Maria}\ \bibnamefont
  {Sessolo}}, \ and\ \bibinfo {author} {\bibfnamefont {Sebastian}\ \bibnamefont
  {Trojanowski}},\ }\bibfield  {title} {\enquote {\bibinfo {title} {{WIMP dark
  matter candidates and searches - current status and future prospects}},}\
  }\href {\doibase 10.1088/1361-6633/aab913} {\bibfield  {journal} {\bibinfo
  {journal} {Reports on Progress in Physics}\ }\textbf {\bibinfo {volume}
  {81}},\ \bibinfo {pages} {066201} (\bibinfo {year} {2018})}\BibitemShut
  {NoStop}%
\bibitem [{\citenamefont {Aprile}\ \emph {et~al.}(2011)\citenamefont {Aprile}
  \emph {et~al.}}]{XENON10}%
  \BibitemOpen
  \bibfield  {author} {\bibinfo {author} {\bibfnamefont {E.}~\bibnamefont
  {Aprile}} \emph {et~al.} (\bibinfo {collaboration} {XENON}),\ }\bibfield
  {title} {\enquote {\bibinfo {title} {{Design and performance of the XENON10
  dark matter experiment}},}\ }\href {\doibase
  10.1016/j.astropartphys.2011.01.006} {\bibfield  {journal} {\bibinfo
  {journal} {Astroparticle Physics}\ }\textbf {\bibinfo {volume} {34}},\
  \bibinfo {pages} {679--698} (\bibinfo {year} {2011})}\BibitemShut {NoStop}%
\bibitem [{\citenamefont {Aprile}\ \emph {et~al.}(2012)\citenamefont {Aprile}
  \emph {et~al.}}]{APRILE2012573_xe100_instr}%
  \BibitemOpen
  \bibfield  {author} {\bibinfo {author} {\bibfnamefont {E.}~\bibnamefont
  {Aprile}} \emph {et~al.},\ }\bibfield  {title} {\enquote {\bibinfo {title}
  {{The XENON100 dark matter experiment}},}\ }\href {\doibase
  10.1016/j.astropartphys.2012.01.003} {\bibfield  {journal} {\bibinfo
  {journal} {Astroparticle Physics}\ }\textbf {\bibinfo {volume} {35}},\
  \bibinfo {pages} {573--590} (\bibinfo {year} {2012})}\BibitemShut {NoStop}%
\bibitem [{\citenamefont {Aprile}\ \emph
  {et~al.}(2017{\natexlab{a}})\citenamefont {Aprile} \emph
  {et~al.}}]{XENON1T_instrument}%
  \BibitemOpen
  \bibfield  {author} {\bibinfo {author} {\bibfnamefont {E.}~\bibnamefont
  {Aprile}} \emph {et~al.} (\bibinfo {collaboration} {XENON}),\ }\bibfield
  {title} {\enquote {\bibinfo {title} {{The XENON1T dark matter experiment}},}\
  }\href {\doibase 10.1140/epjc/s10052-017-5326-3} {\bibfield  {journal}
  {\bibinfo  {journal} {Eur. Phys. J.}\ }\textbf {\bibinfo {volume} {C77}},\
  \bibinfo {pages} {881} (\bibinfo {year} {2017}{\natexlab{a}})},\ \Eprint
  {http://arxiv.org/abs/1708.07051} {arXiv:1708.07051 [astro-ph.IM]}
  \BibitemShut {NoStop}%
\bibitem [{\citenamefont {Alner}\ \emph {et~al.}(2007)\citenamefont {Alner}
  \emph {et~al.}}]{ALNER2007287_zeplinII}%
  \BibitemOpen
  \bibfield  {author} {\bibinfo {author} {\bibfnamefont {G.J.}\ \bibnamefont
  {Alner}} \emph {et~al.},\ }\bibfield  {title} {\enquote {\bibinfo {title}
  {{First limits on WIMP nuclear recoil signals in ZEPLIN-II: A two-phase xenon
  detector for dark matter detection}},}\ }\href {\doibase
  10.1016/j.astropartphys.2007.06.002} {\bibfield  {journal} {\bibinfo
  {journal} {Astroparticle Physics}\ }\textbf {\bibinfo {volume} {28}},\
  \bibinfo {pages} {287--302} (\bibinfo {year} {2007})}\BibitemShut {NoStop}%
\bibitem [{\citenamefont {Akimov}\ \emph {et~al.}(2007)\citenamefont {Akimov}
  \emph {et~al.}}]{AKIMOV200746_zeplinIII}%
  \BibitemOpen
  \bibfield  {author} {\bibinfo {author} {\bibfnamefont {D.Yu.}\ \bibnamefont
  {Akimov}} \emph {et~al.},\ }\bibfield  {title} {\enquote {\bibinfo {title}
  {{The ZEPLIN-III dark matter detector: Instrument design, manufacture and
  commissioning}},}\ }\href {\doibase 10.1016/j.astropartphys.2006.09.005}
  {\bibfield  {journal} {\bibinfo  {journal} {Astroparticle Physics}\ }\textbf
  {\bibinfo {volume} {27}},\ \bibinfo {pages} {46--60} (\bibinfo {year}
  {2007})}\BibitemShut {NoStop}%
\bibitem [{\citenamefont {Akerib}\ \emph {et~al.}(2013)\citenamefont {Akerib}
  \emph {et~al.}}]{AKERIB2013111_LUX_instr}%
  \BibitemOpen
  \bibfield  {author} {\bibinfo {author} {\bibfnamefont {D.S.}\ \bibnamefont
  {Akerib}} \emph {et~al.},\ }\bibfield  {title} {\enquote {\bibinfo {title}
  {{The Large Underground Xenon (LUX) experiment}},}\ }\href {\doibase
  10.1016/j.nima.2012.11.135} {\bibfield  {journal} {\bibinfo  {journal}
  {Nuclear Instruments and Methods in Physics Research Section A: Accelerators,
  Spectrometers, Detectors and Associated Equipment}\ }\textbf {\bibinfo
  {volume} {704}},\ \bibinfo {pages} {111--126} (\bibinfo {year}
  {2013})}\BibitemShut {NoStop}%
\bibitem [{\citenamefont {Baudis}(2023)}]{Baudis:2023pzu}%
  \BibitemOpen
  \bibfield  {author} {\bibinfo {author} {\bibfnamefont {Laura}\ \bibnamefont
  {Baudis}},\ }\bibfield  {title} {\enquote {\bibinfo {title} {{Dual-phase
  xenon time projection chambers for~rare-event searches}},}\ }\href {\doibase
  10.1098/rsta.2023.0083} {\bibfield  {journal} {\bibinfo  {journal} {Phil.
  Trans. Roy. Soc. Lond. A}\ }\textbf {\bibinfo {volume} {382}},\ \bibinfo
  {pages} {20230083} (\bibinfo {year} {2023})},\ \Eprint
  {http://arxiv.org/abs/2311.05320} {arXiv:2311.05320 [physics.ins-det]}
  \BibitemShut {NoStop}%
\bibitem [{\citenamefont {Aalseth}\ \emph {et~al.}(2018)\citenamefont {Aalseth}
  \emph {et~al.}}]{Darkside20k_2018}%
  \BibitemOpen
  \bibfield  {author} {\bibinfo {author} {\bibfnamefont {C.E}\ \bibnamefont
  {Aalseth}} \emph {et~al.},\ }\bibfield  {title} {\enquote {\bibinfo {title}
  {{DarkSide-20k: a 20 Tonne two-phase LAr TPC for direct dark matter detection
  at LNGS}},}\ }\href {\doibase 10.1140/epjp/i2018-11973-4} {\bibfield
  {journal} {\bibinfo  {journal} {Eur. Phys. J. Plus}\ }\textbf {\bibinfo
  {volume} {133}},\ \bibinfo {pages} {131} (\bibinfo {year}
  {2018})}\BibitemShut {NoStop}%
\bibitem [{\citenamefont {Agostini}\ \emph {et~al.}(2023)\citenamefont
  {Agostini} \emph {et~al.}}]{RevModPhys_95_025002_NDBD}%
  \BibitemOpen
  \bibfield  {author} {\bibinfo {author} {\bibfnamefont {Matteo}\ \bibnamefont
  {Agostini}} \emph {et~al.},\ }\bibfield  {title} {\enquote {\bibinfo {title}
  {{Toward the discovery of matter creation with neutrinoless
  $\ensuremath{\beta}\ensuremath{\beta}$ decay}},}\ }\href {\doibase
  10.1103/RevModPhys.95.025002} {\bibfield  {journal} {\bibinfo  {journal}
  {Rev. Mod. Phys.}\ }\textbf {\bibinfo {volume} {95}},\ \bibinfo {pages}
  {025002} (\bibinfo {year} {2023})}\BibitemShut {NoStop}%
\bibitem [{\citenamefont {Mart\'\i{}n-Albo}\ \emph {et~al.}(2016)\citenamefont
  {Mart\'\i{}n-Albo} \emph {et~al.}}]{NEXT:2015wlq}%
  \BibitemOpen
  \bibfield  {author} {\bibinfo {author} {\bibfnamefont {J.}~\bibnamefont
  {Mart\'\i{}n-Albo}} \emph {et~al.} (\bibinfo {collaboration} {NEXT}),\
  }\bibfield  {title} {\enquote {\bibinfo {title} {{Sensitivity of NEXT-100 to
  Neutrinoless Double Beta Decay}},}\ }\href {\doibase 10.1007/JHEP05(2016)159}
  {\bibfield  {journal} {\bibinfo  {journal} {JHEP}\ }\textbf {\bibinfo
  {volume} {05}},\ \bibinfo {pages} {159} (\bibinfo {year} {2016})},\ \Eprint
  {http://arxiv.org/abs/1511.09246} {arXiv:1511.09246 [physics.ins-det]}
  \BibitemShut {NoStop}%
\bibitem [{\citenamefont {Auger}\ \emph {et~al.}(2012)\citenamefont {Auger}
  \emph {et~al.}}]{M_Auger_2012_exo200_instr}%
  \BibitemOpen
  \bibfield  {author} {\bibinfo {author} {\bibfnamefont {M}~\bibnamefont
  {Auger}} \emph {et~al.},\ }\bibfield  {title} {\enquote {\bibinfo {title}
  {{The EXO-200 detector, part I: detector design and construction}},}\ }\href
  {\doibase 10.1088/1748-0221/7/05/P05010} {\bibfield  {journal} {\bibinfo
  {journal} {Journal of Instrumentation}\ }\textbf {\bibinfo {volume} {7}},\
  \bibinfo {pages} {P05010} (\bibinfo {year} {2012})}\BibitemShut {NoStop}%
\bibitem [{\citenamefont {Adhikari}\ \emph {et~al.}(2021)\citenamefont
  {Adhikari} \emph {et~al.}}]{nEXO_Rn222}%
  \BibitemOpen
  \bibfield  {author} {\bibinfo {author} {\bibfnamefont {G.}~\bibnamefont
  {Adhikari}} \emph {et~al.},\ }\bibfield  {title} {\enquote {\bibinfo {title}
  {{nEXO: neutrinoless double beta decay search beyond 10$^{28}$ year half-life
  sensitivity}},}\ }\href {\doibase 10.1088/1361-6471/ac3631} {\bibfield
  {journal} {\bibinfo  {journal} {Journal of Physics G: Nuclear and Particle
  Physics}\ }\textbf {\bibinfo {volume} {49}},\ \bibinfo {pages} {015104}
  (\bibinfo {year} {2021})}\BibitemShut {NoStop}%
\bibitem [{\citenamefont {Aprile}\ \emph {et~al.}(2019)\citenamefont {Aprile}
  \emph {et~al.}}]{XENON1T_ECEC_nature}%
  \BibitemOpen
  \bibfield  {author} {\bibinfo {author} {\bibfnamefont {E.}~\bibnamefont
  {Aprile}} \emph {et~al.} (\bibinfo {collaboration} {XENON}),\ }\bibfield
  {title} {\enquote {\bibinfo {title} {{Observation of two-neutrino double
  electron capture in $^{124}$Xe with XENON1T}},}\ }\href {\doibase
  10.1038/s41586-019-1124-4} {\bibfield  {journal} {\bibinfo  {journal}
  {Nature}\ }\textbf {\bibinfo {volume} {568}},\ \bibinfo {pages} {532--535}
  (\bibinfo {year} {2019})},\ \Eprint {http://arxiv.org/abs/1904.11002}
  {arXiv:1904.11002 [nucl-ex]} \BibitemShut {NoStop}%
\bibitem [{\citenamefont {Aprile}\ \emph
  {et~al.}(2022{\natexlab{b}})\citenamefont {Aprile} \emph
  {et~al.}}]{XENON1T_ECEC0nbb_prc}%
  \BibitemOpen
  \bibfield  {author} {\bibinfo {author} {\bibfnamefont {E.}~\bibnamefont
  {Aprile}} \emph {et~al.} (\bibinfo {collaboration} {XENON}),\ }\bibfield
  {title} {\enquote {\bibinfo {title} {{Double-Weak Decays of $^{124}$Xe and
  $^{136}$Xe in the XENON1T and XENONnT Experiments}},}\ }\href {\doibase
  10.1103/PhysRevC.106.024328} {\bibfield  {journal} {\bibinfo  {journal}
  {Phys. Rev. C}\ }\textbf {\bibinfo {volume} {106}},\ \bibinfo {pages}
  {024328} (\bibinfo {year} {2022}{\natexlab{b}})},\ \Eprint
  {http://arxiv.org/abs/2205.04158} {arXiv:2205.04158 [hep-ex]} \BibitemShut
  {NoStop}%
\bibitem [{\citenamefont {Aprile}\ \emph
  {et~al.}(2017{\natexlab{b}})\citenamefont {Aprile} \emph
  {et~al.}}]{XENON1T_kr_removal}%
  \BibitemOpen
  \bibfield  {author} {\bibinfo {author} {\bibfnamefont {E.}~\bibnamefont
  {Aprile}} \emph {et~al.} (\bibinfo {collaboration} {XENON}),\ }\bibfield
  {title} {\enquote {\bibinfo {title} {{Removing krypton from xenon by
  cryogenic distillation to the ppq level}},}\ }\href {\doibase
  10.1140/epjc/s10052-017-4757-1} {\bibfield  {journal} {\bibinfo  {journal}
  {Eur. Phys. J. C}\ }\textbf {\bibinfo {volume} {77}},\ \bibinfo {pages} {275}
  (\bibinfo {year} {2017}{\natexlab{b}})},\ \Eprint
  {http://arxiv.org/abs/1612.04284} {arXiv:1612.04284 [physics.ins-det]}
  \BibitemShut {NoStop}%
\bibitem [{\citenamefont {{ENSDF database}}(2023)}]{ENSDF}%
  \BibitemOpen
  \bibfield  {author} {\bibinfo {author} {\bibnamefont {{ENSDF database}}},\
  }\href {\doibase 10.18139/nndc.ensdf/1845010} {\enquote {\bibinfo {title}
  {{http://www.nndc.bnl.gov/ensarchivals/}},}\ } (\bibinfo {year}
  {2023})\BibitemShut {NoStop}%
\bibitem [{\citenamefont {Aprile}\ \emph
  {et~al.}(2024{\natexlab{b}})\citenamefont {Aprile} \emph
  {et~al.}}]{XENONCollaborationP:2024xwn}%
  \BibitemOpen
  \bibfield  {author} {\bibinfo {author} {\bibfnamefont {E.}~\bibnamefont
  {Aprile}} \emph {et~al.} (\bibinfo {collaboration} {(XENON
  Collaboration)\textparagraph{}, XENON}),\ }\bibfield  {title} {\enquote
  {\bibinfo {title} {{Offline tagging of radon-induced backgrounds in XENON1T
  and applicability to other liquid xenon time projection chambers}},}\ }\href
  {\doibase 10.1103/PhysRevD.110.012011} {\bibfield  {journal} {\bibinfo
  {journal} {Phys. Rev. D}\ }\textbf {\bibinfo {volume} {110}},\ \bibinfo
  {pages} {012011} (\bibinfo {year} {2024}{\natexlab{b}})},\ \Eprint
  {http://arxiv.org/abs/2403.14878} {arXiv:2403.14878 [hep-ex]} \BibitemShut
  {NoStop}%
\bibitem [{\citenamefont {Aalbers}\ \emph
  {et~al.}(2024{\natexlab{a}})\citenamefont {Aalbers} \emph
  {et~al.}}]{LZCollaboration:2024lux}%
  \BibitemOpen
  \bibfield  {author} {\bibinfo {author} {\bibfnamefont {J.}~\bibnamefont
  {Aalbers}} \emph {et~al.} (\bibinfo {collaboration} {LZ Collaboration}),\
  }\bibfield  {title} {\enquote {\bibinfo {title} {{Dark Matter Search Results
  from 4.2 Tonne-Years of Exposure of the LUX-ZEPLIN (LZ) Experiment}},}\
  }\href@noop {} {\  (\bibinfo {year} {2024}{\natexlab{a}})},\ \Eprint
  {http://arxiv.org/abs/2410.17036} {arXiv:2410.17036 [hep-ex]} \BibitemShut
  {NoStop}%
\bibitem [{\citenamefont {Brown}\ \emph {et~al.}(2018)\citenamefont {Brown}
  \emph {et~al.}}]{Xe1T_magpump}%
  \BibitemOpen
  \bibfield  {author} {\bibinfo {author} {\bibfnamefont {Ethan}\ \bibnamefont
  {Brown}} \emph {et~al.},\ }\bibfield  {title} {\enquote {\bibinfo {title}
  {{Magnetically-coupled piston pump for high-purity gas applications}},}\
  }\href {\doibase 10.1140/epjc/s10052-018-6062-z} {\bibfield  {journal}
  {\bibinfo  {journal} {Eur. Phys. J. C}\ }\textbf {\bibinfo {volume} {78}},\
  \bibinfo {pages} {604} (\bibinfo {year} {2018})},\ \Eprint
  {http://arxiv.org/abs/1803.08498} {arXiv:1803.08498 [physics.ins-det]}
  \BibitemShut {NoStop}%
\bibitem [{\citenamefont {Aprile}\ \emph {et~al.}(2021)\citenamefont {Aprile}
  \emph {et~al.}}]{XENON1T_radon_emanation}%
  \BibitemOpen
  \bibfield  {author} {\bibinfo {author} {\bibfnamefont {E.}~\bibnamefont
  {Aprile}} \emph {et~al.} (\bibinfo {collaboration} {XENON}),\ }\bibfield
  {title} {\enquote {\bibinfo {title} {{$^{222}$Rn emanation measurements for
  the XENON1T experiment}},}\ }\href {\doibase 10.1140/epjc/s10052-020-08777-z}
  {\bibfield  {journal} {\bibinfo  {journal} {Eur. Phys. J. C}\ }\textbf
  {\bibinfo {volume} {81}},\ \bibinfo {pages} {337} (\bibinfo {year} {2021})},\
  \Eprint {http://arxiv.org/abs/2009.13981} {arXiv:2009.13981
  [physics.ins-det]} \BibitemShut {NoStop}%
\bibitem [{\citenamefont {Aalbers}\ \emph
  {et~al.}(2023{\natexlab{b}})\citenamefont {Aalbers} \emph
  {et~al.}}]{DARWIN_XLZD_White_Paper}%
  \BibitemOpen
  \bibfield  {author} {\bibinfo {author} {\bibfnamefont {J.}~\bibnamefont
  {Aalbers}} \emph {et~al.},\ }\bibfield  {title} {\enquote {\bibinfo {title}
  {{A next-generation liquid xenon observatory for dark matter and neutrino
  physics}},}\ }\href {\doibase 10.1088/1361-6471/ac841a} {\bibfield  {journal}
  {\bibinfo  {journal} {J. Phys. G}\ }\textbf {\bibinfo {volume} {50}},\
  \bibinfo {pages} {013001} (\bibinfo {year} {2023}{\natexlab{b}})},\ \Eprint
  {http://arxiv.org/abs/2203.02309} {arXiv:2203.02309 [physics.ins-det]}
  \BibitemShut {NoStop}%
\bibitem [{\citenamefont {Plante}\ \emph {et~al.}(2022)\citenamefont {Plante},
  \citenamefont {Aprile}, \citenamefont {Howlett},\ and\ \citenamefont
  {Zhang}}]{Plante:2022khm}%
  \BibitemOpen
  \bibfield  {author} {\bibinfo {author} {\bibfnamefont {G.}~\bibnamefont
  {Plante}}, \bibinfo {author} {\bibfnamefont {E.}~\bibnamefont {Aprile}},
  \bibinfo {author} {\bibfnamefont {J.}~\bibnamefont {Howlett}}, \ and\
  \bibinfo {author} {\bibfnamefont {Y.}~\bibnamefont {Zhang}},\ }\bibfield
  {title} {\enquote {\bibinfo {title} {{Liquid-phase purification for
  multi-tonne xenon detectors}},}\ }\href {\doibase
  10.1140/epjc/s10052-022-10832-w} {\bibfield  {journal} {\bibinfo  {journal}
  {Eur. Phys. J. C}\ }\textbf {\bibinfo {volume} {82}},\ \bibinfo {pages} {860}
  (\bibinfo {year} {2022})},\ \Eprint {http://arxiv.org/abs/2205.07336}
  {arXiv:2205.07336 [physics.ins-det]} \BibitemShut {NoStop}%
\bibitem [{\citenamefont {Aprile}\ \emph
  {et~al.}(2022{\natexlab{c}})\citenamefont {Aprile} \emph
  {et~al.}}]{radioactivity_XENONnT}%
  \BibitemOpen
  \bibfield  {author} {\bibinfo {author} {\bibfnamefont {E.}~\bibnamefont
  {Aprile}} \emph {et~al.} (\bibinfo {collaboration} {XENON}),\ }\bibfield
  {title} {\enquote {\bibinfo {title} {{Material radiopurity control in the
  XENONnT experiment}},}\ }\href {\doibase 10.1140/epjc/s10052-022-10345-6}
  {\bibfield  {journal} {\bibinfo  {journal} {Eur. Phys. J. C}\ }\textbf
  {\bibinfo {volume} {82}},\ \bibinfo {pages} {599} (\bibinfo {year}
  {2022}{\natexlab{c}})},\ \Eprint {http://arxiv.org/abs/2112.05629}
  {arXiv:2112.05629 [physics.ins-det]} \BibitemShut {NoStop}%
\bibitem [{\citenamefont {Murra}\ \emph
  {et~al.}(2022{\natexlab{a}})\citenamefont {Murra}, \citenamefont {Schulte},
  \citenamefont {Huhmann},\ and\ \citenamefont {Weinheimer}}]{RAD_EPJC}%
  \BibitemOpen
  \bibfield  {author} {\bibinfo {author} {\bibfnamefont {M.}~\bibnamefont
  {Murra}}, \bibinfo {author} {\bibfnamefont {D.}~\bibnamefont {Schulte}},
  \bibinfo {author} {\bibfnamefont {C.}~\bibnamefont {Huhmann}}, \ and\
  \bibinfo {author} {\bibfnamefont {C.}~\bibnamefont {Weinheimer}},\ }\bibfield
   {title} {\enquote {\bibinfo {title} {{Design, construction and commissioning
  of a high-flow radon removal system for XENONnT}},}\ }\href {\doibase
  10.1140/epjc/s10052-022-11001-9} {\bibfield  {journal} {\bibinfo  {journal}
  {Eur. Phys. J. C}\ }\textbf {\bibinfo {volume} {82}},\ \bibinfo {pages}
  {1104} (\bibinfo {year} {2022}{\natexlab{a}})},\ \Eprint
  {http://arxiv.org/abs/2205.11492} {arXiv:2205.11492 [physics.ins-det]}
  \BibitemShut {NoStop}%
\bibitem [{\citenamefont {Lemmon}\ \emph {et~al.}(2024)\citenamefont {Lemmon},
  \citenamefont {Bell}, \citenamefont {Huber},\ and\ \citenamefont
  {McLinden}}]{NIST}%
  \BibitemOpen
  \bibfield  {author} {\bibinfo {author} {\bibfnamefont {Eric~W.}\ \bibnamefont
  {Lemmon}}, \bibinfo {author} {\bibfnamefont {Ian~H.}\ \bibnamefont {Bell}},
  \bibinfo {author} {\bibfnamefont {Marcia~L.}\ \bibnamefont {Huber}}, \ and\
  \bibinfo {author} {\bibfnamefont {Mark~O.}\ \bibnamefont {McLinden}},\
  }\href@noop {} {\emph {\bibinfo {title} {{Thermophysical Properties of Fluid
  Systems}}}}\ (\bibinfo  {publisher} {{NIST Chemistry WebBook, NIST Standard
  Reference Database Number 69, Eds. P.J. Linstrom and W.G. Mallard}},\
  \bibinfo {year} {2024})\BibitemShut {NoStop}%
\bibitem [{\citenamefont {Murra}\ \emph
  {et~al.}(2022{\natexlab{b}})\citenamefont {Murra} \emph
  {et~al.}}]{RAD_HE_JINST}%
  \BibitemOpen
  \bibfield  {author} {\bibinfo {author} {\bibfnamefont {M.}~\bibnamefont
  {Murra}} \emph {et~al.},\ }\bibfield  {title} {\enquote {\bibinfo {title}
  {{Cryogenic bath-type heat exchangers for ultra-pure noble gas
  applications}},}\ }\href {\doibase 10.1088/1748-0221/17/05/P05037} {\bibfield
   {journal} {\bibinfo  {journal} {JINST}\ }\textbf {\bibinfo {volume} {17}},\
  \bibinfo {pages} {P05037} (\bibinfo {year} {2022}{\natexlab{b}})},\ \Eprint
  {http://arxiv.org/abs/2203.01026} {arXiv:2203.01026 [physics.ins-det]}
  \BibitemShut {NoStop}%
\bibitem [{\citenamefont {Schulte}\ \emph {et~al.}(2021)\citenamefont {Schulte}
  \emph {et~al.}}]{RAD_4MP_JINST}%
  \BibitemOpen
  \bibfield  {author} {\bibinfo {author} {\bibfnamefont {D.}~\bibnamefont
  {Schulte}} \emph {et~al.},\ }\bibfield  {title} {\enquote {\bibinfo {title}
  {{Ultra-clean radon-free four cylinder magnetically-coupled piston pump}},}\
  }\href {\doibase 10.1088/1748-0221/16/09/P09011} {\bibfield  {journal}
  {\bibinfo  {journal} {JINST}\ }\textbf {\bibinfo {volume} {16}},\ \bibinfo
  {pages} {P09011} (\bibinfo {year} {2021})},\ \Eprint
  {http://arxiv.org/abs/2107.00755} {arXiv:2107.00755 [physics.ins-det]}
  \BibitemShut {NoStop}%
\bibitem [{\citenamefont {B\'e}\ \emph {et~al.}(2008)\citenamefont {B\'e},
  \citenamefont {Chist\'e}, \citenamefont {Dulieu}, \citenamefont {Browne},
  \citenamefont {Chechev}, \citenamefont {Kuzmenko}, \citenamefont {Kondev},
  \citenamefont {Luca}, \citenamefont {Gal\'an}, \citenamefont {Pearce},\ and\
  \citenamefont {Huang}}]{TabRad_v4}%
  \BibitemOpen
  \bibfield  {author} {\bibinfo {author} {\bibfnamefont {M.-M.}\ \bibnamefont
  {B\'e}}, \bibinfo {author} {\bibfnamefont {V.}~\bibnamefont {Chist\'e}},
  \bibinfo {author} {\bibfnamefont {C.}~\bibnamefont {Dulieu}}, \bibinfo
  {author} {\bibfnamefont {E.}~\bibnamefont {Browne}}, \bibinfo {author}
  {\bibfnamefont {V.}~\bibnamefont {Chechev}}, \bibinfo {author} {\bibfnamefont
  {N.}~\bibnamefont {Kuzmenko}}, \bibinfo {author} {\bibfnamefont
  {F.}~\bibnamefont {Kondev}}, \bibinfo {author} {\bibfnamefont
  {A.}~\bibnamefont {Luca}}, \bibinfo {author} {\bibfnamefont {M.}~\bibnamefont
  {Gal\'an}}, \bibinfo {author} {\bibfnamefont {A.}~\bibnamefont {Pearce}}, \
  and\ \bibinfo {author} {\bibfnamefont {X.}~\bibnamefont {Huang}},\ }\href
  {http://www.bipm.org/utils/common/pdf/monographieRI/Monographie_BIPM-5_Tables_Vol4.pdf}
  {\emph {\bibinfo {title} {Table of Radionuclides}}},\ \bibinfo {series}
  {Monographie BIPM-5}, Vol.~\bibinfo {volume} {4}\ (\bibinfo  {publisher}
  {Bureau International des Poids et Mesures},\ \bibinfo {address} {Pavillon de
  Breteuil, F-92310 S\`evres, France},\ \bibinfo {year} {2008})\BibitemShut
  {NoStop}%
\bibitem [{\citenamefont {{Ziegler}}\ \emph {et~al.}(2010)\citenamefont
  {{Ziegler}}, \citenamefont {{Ziegler}},\ and\ \citenamefont
  {{Biersack}}}]{Ziegler:2010}%
  \BibitemOpen
  \bibfield  {author} {\bibinfo {author} {\bibfnamefont {J.~F.}\ \bibnamefont
  {{Ziegler}}}, \bibinfo {author} {\bibfnamefont {M.~D.}\ \bibnamefont
  {{Ziegler}}}, \ and\ \bibinfo {author} {\bibfnamefont {J.~P.}\ \bibnamefont
  {{Biersack}}},\ }\bibfield  {title} {\enquote {\bibinfo {title} {{SRIM - The
  stopping and range of ions in matter (2010)}},}\ }\href {\doibase
  10.1016/j.nimb.2010.02.091} {\bibfield  {journal} {\bibinfo  {journal} {Nucl.
  Instrum. Meth.}\ }\textbf {\bibinfo {volume} {B268}},\ \bibinfo {pages}
  {1818} (\bibinfo {year} {2010})}\BibitemShut {NoStop}%
\bibitem [{\citenamefont {Aprile}\ \emph
  {et~al.}(2024{\natexlab{c}})\citenamefont {Aprile} \emph
  {et~al.}}]{XENONnT_analys1}%
  \BibitemOpen
  \bibfield  {author} {\bibinfo {author} {\bibfnamefont {E.}~\bibnamefont
  {Aprile}} \emph {et~al.} (\bibinfo {collaboration} {XENON}),\ }\bibfield
  {title} {\enquote {\bibinfo {title} {{XENONnT Analysis: Signal
  Reconstruction, Calibration and Event Selection}},}\ }\href@noop {}
  {\bibfield  {journal} {\bibinfo  {journal} {Phys. Rev. D}\ }\textbf {\bibinfo
  {volume} {\text{accepted}}} (\bibinfo {year} {2024}{\natexlab{c}})},\ \Eprint
  {http://arxiv.org/abs/2409.08778} {arXiv:2409.08778 [hep-ex]} \BibitemShut
  {NoStop}%
\bibitem [{\citenamefont {Aprile}\ \emph
  {et~al.}(2024{\natexlab{d}})\citenamefont {Aprile} \emph
  {et~al.}}]{XENON_B8}%
  \BibitemOpen
  \bibfield  {author} {\bibinfo {author} {\bibfnamefont {E.}~\bibnamefont
  {Aprile}} \emph {et~al.} (\bibinfo {collaboration} {XENON}),\ }\bibfield
  {title} {\enquote {\bibinfo {title} {{First Indication of Solar $^8$B
  Neutrinos via Coherent Elastic Neutrino-Nucleus Scattering with XENONnT}},}\
  }\href {\doibase 10.1103/PhysRevLett.133.191002} {\bibfield  {journal}
  {\bibinfo  {journal} {Phys. Rev. Lett.}\ }\textbf {\bibinfo {volume} {133}},\
  \bibinfo {pages} {191002} (\bibinfo {year} {2024}{\natexlab{d}})},\ \Eprint
  {http://arxiv.org/abs/2408.02877} {arXiv:2408.02877 [hep-ex]} \BibitemShut
  {NoStop}%
\bibitem [{\citenamefont {Guida}(2021)}]{Guida:2021elc}%
  \BibitemOpen
  \bibfield  {author} {\bibinfo {author} {\bibfnamefont {Matteo}\ \bibnamefont
  {Guida}},\ }\emph {\bibinfo {title} {{Position Reconstruction Based on Prompt
  Scintillation Light in XENONnT Exploiting Deep Learning}}},\ \href {\doibase
  10.5281/zenodo.6347581} {Master's thesis},\ \bibinfo  {school} {Padua U.}
  (\bibinfo {year} {2021})\BibitemShut {NoStop}%
\bibitem [{\citenamefont {Aalbers}\ \emph
  {et~al.}(2024{\natexlab{b}})\citenamefont {Aalbers} \emph
  {et~al.}}]{XLZD:2024gxx}%
  \BibitemOpen
  \bibfield  {author} {\bibinfo {author} {\bibfnamefont {J.}~\bibnamefont
  {Aalbers}} \emph {et~al.} (\bibinfo {collaboration} {XLZD}),\ }\bibfield
  {title} {\enquote {\bibinfo {title} {{The XLZD Design Book: Towards the
  Next-Generation Liquid Xenon Observatory for Dark Matter and Neutrino
  Physics}},}\ }\href@noop {} {\  (\bibinfo {year} {2024}{\natexlab{b}})},\
  \Eprint {http://arxiv.org/abs/2410.17137} {arXiv:2410.17137 [hep-ex]}
  \BibitemShut {NoStop}%
\bibitem [{\citenamefont {Aprile}\ \emph
  {et~al.}(2017{\natexlab{c}})\citenamefont {Aprile} \emph
  {et~al.}}]{Aprile:2017radon_xe100}%
  \BibitemOpen
  \bibfield  {author} {\bibinfo {author} {\bibfnamefont {E.}~\bibnamefont
  {Aprile}} \emph {et~al.} (\bibinfo {collaboration} {XENON100}),\ }\bibfield
  {title} {\enquote {\bibinfo {title} {{Online $^{222}$Rn removal by cryogenic
  distillation in the XENON100 experiment}},}\ }\href {\doibase
  10.1140/epjc/s10052-017-4902-x} {\bibfield  {journal} {\bibinfo  {journal}
  {Eur. Phys. J.}\ }\textbf {\bibinfo {volume} {C77}},\ \bibinfo {pages} {358}
  (\bibinfo {year} {2017}{\natexlab{c}})},\ \Eprint
  {http://arxiv.org/abs/1702.06942} {arXiv:1702.06942 [physics.ins-det]}
  \BibitemShut {NoStop}%
\bibitem [{\citenamefont {Albert}\ \emph {et~al.}(2015)\citenamefont {Albert}
  \emph {et~al.}}]{EXO200_Rn222}%
  \BibitemOpen
  \bibfield  {author} {\bibinfo {author} {\bibfnamefont {J.~B.}\ \bibnamefont
  {Albert}} \emph {et~al.} (\bibinfo {collaboration} {EXO-200 Collaboration}),\
  }\bibfield  {title} {\enquote {\bibinfo {title} {{Investigation of
  radioactivity-induced backgrounds in EXO-200}},}\ }\href {\doibase
  10.1103/PhysRevC.92.015503} {\bibfield  {journal} {\bibinfo  {journal} {Phys.
  Rev. C}\ }\textbf {\bibinfo {volume} {92}},\ \bibinfo {pages} {015503}
  (\bibinfo {year} {2015})}\BibitemShut {NoStop}%
\bibitem [{\citenamefont {Bradley}\ \emph {et~al.}(2015)\citenamefont {Bradley}
  \emph {et~al.}}]{LUX_Rn222}%
  \BibitemOpen
  \bibfield  {author} {\bibinfo {author} {\bibfnamefont {A.}~\bibnamefont
  {Bradley}} \emph {et~al.},\ }\bibfield  {title} {\enquote {\bibinfo {title}
  {{Radon-related Backgrounds in the LUX Dark Matter Search}},}\ }\href
  {\doibase 10.1016/j.phpro.2014.12.067} {\bibfield  {journal} {\bibinfo
  {journal} {Physics Procedia}\ }\textbf {\bibinfo {volume} {61}},\ \bibinfo
  {pages} {658--665} (\bibinfo {year} {2015})},\ \bibinfo {note} {13th
  International Conference on Topics in Astroparticle and Underground Physics,
  TAUP 2013}\BibitemShut {NoStop}%
\bibitem [{\citenamefont {Tan}\ \emph {et~al.}(2016)\citenamefont {Tan} \emph
  {et~al.}}]{PandaX_II_Rn222}%
  \BibitemOpen
  \bibfield  {author} {\bibinfo {author} {\bibfnamefont {A.}~\bibnamefont
  {Tan}} \emph {et~al.} (\bibinfo {collaboration} {PandaX-II Collaboration}),\
  }\bibfield  {title} {\enquote {\bibinfo {title} {{Dark Matter Results from
  First 98.7 Days of Data from the PandaX-II Experiment}},}\ }\href {\doibase
  10.1103/PhysRevLett.117.121303} {\bibfield  {journal} {\bibinfo  {journal}
  {Phys. Rev. Lett.}\ }\textbf {\bibinfo {volume} {117}},\ \bibinfo {pages}
  {121303} (\bibinfo {year} {2016})}\BibitemShut {NoStop}%
\bibitem [{\citenamefont {Abe}\ \emph {et~al.}(2013)\citenamefont {Abe} \emph
  {et~al.}}]{XMASS_Rn222}%
  \BibitemOpen
  \bibfield  {author} {\bibinfo {author} {\bibfnamefont {K.}~\bibnamefont
  {Abe}} \emph {et~al.},\ }\bibfield  {title} {\enquote {\bibinfo {title}
  {{XMASS detector}},}\ }\href {\doibase 10.1016/j.nima.2013.03.059} {\bibfield
   {journal} {\bibinfo  {journal} {Nucl. Instrum. Methods Phys. Res. A}\
  }\textbf {\bibinfo {volume} {716}},\ \bibinfo {pages} {78--85} (\bibinfo
  {year} {2013})}\BibitemShut {NoStop}%
\bibitem [{\citenamefont {Aalbers}\ \emph
  {et~al.}(2023{\natexlab{c}})\citenamefont {Aalbers} \emph
  {et~al.}}]{LZ_Rn222}%
  \BibitemOpen
  \bibfield  {author} {\bibinfo {author} {\bibfnamefont {J.}~\bibnamefont
  {Aalbers}} \emph {et~al.} (\bibinfo {collaboration} {The LUX-ZEPLIN
  Collaboration}),\ }\bibfield  {title} {\enquote {\bibinfo {title}
  {{Background determination for the LUX-ZEPLIN dark matter experiment}},}\
  }\href {\doibase 10.1103/PhysRevD.108.012010} {\bibfield  {journal} {\bibinfo
   {journal} {Phys. Rev. D}\ }\textbf {\bibinfo {volume} {108}},\ \bibinfo
  {pages} {012010} (\bibinfo {year} {2023}{\natexlab{c}})}\BibitemShut
  {NoStop}%
\bibitem [{\citenamefont {Abdukerim}\ \emph {et~al.}(2025)\citenamefont
  {Abdukerim} \emph {et~al.}}]{PandaX:2024oxq}%
  \BibitemOpen
  \bibfield  {author} {\bibinfo {author} {\bibfnamefont {Abdusalam}\
  \bibnamefont {Abdukerim}} \emph {et~al.} (\bibinfo {collaboration}
  {PandaX}),\ }\bibfield  {title} {\enquote {\bibinfo {title} {{PandaX-xT: a
  Multi-ten-tonne Liquid Xenon Observatory at the China Jinping Underground
  Laboratory}},}\ }\href {\doibase doi.org/10.1007/s11433-024-2539-y}
  {\bibfield  {journal} {\bibinfo  {journal} {Sci. China Phys. Mech. Astron.}\
  }\textbf {\bibinfo {volume} {68}},\ \bibinfo {pages} {221011} (\bibinfo
  {year} {2025})},\ \Eprint {http://arxiv.org/abs/2402.03596} {arXiv:2402.03596
  [hep-ex]} \BibitemShut {NoStop}%
\bibitem [{\citenamefont {Chen}\ \emph {et~al.}(2017)\citenamefont {Chen} \emph
  {et~al.}}]{CHEN2017656_RRPA}%
  \BibitemOpen
  \bibfield  {author} {\bibinfo {author} {\bibfnamefont {J.-W.}\ \bibnamefont
  {Chen}} \emph {et~al.},\ }\bibfield  {title} {\enquote {\bibinfo {title}
  {{Low-energy electronic recoil in xenon detectors by solar neutrinos}},}\
  }\href {\doibase 10.1016/j.physletb.2017.10.029} {\bibfield  {journal}
  {\bibinfo  {journal} {Physics Letters B}\ }\textbf {\bibinfo {volume}
  {774}},\ \bibinfo {pages} {656--661} (\bibinfo {year} {2017})}\BibitemShut
  {NoStop}%
\bibitem [{\citenamefont {Jörg}\ \emph {et~al.}(2023)\citenamefont {Jörg}
  \emph {et~al.}}]{MPIK_Rn222_source}%
  \BibitemOpen
  \bibfield  {author} {\bibinfo {author} {\bibfnamefont {F.}~\bibnamefont
  {Jörg}} \emph {et~al.},\ }\bibfield  {title} {\enquote {\bibinfo {title}
  {{Production and characterization of a $^{222}$Rn-emanating stainless steel
  source}},}\ }\href {\doibase 10.1016/j.apradiso.2023.110666} {\bibfield
  {journal} {\bibinfo  {journal} {Applied Radiation and Isotopes}\ }\textbf
  {\bibinfo {volume} {194}},\ \bibinfo {pages} {110666} (\bibinfo {year}
  {2023})}\BibitemShut {NoStop}%
\bibitem [{\citenamefont {Agostini}\ \emph {et~al.}(2020)\citenamefont
  {Agostini} \emph {et~al.}}]{solar_neutrino_borexino}%
  \BibitemOpen
  \bibfield  {author} {\bibinfo {author} {\bibfnamefont {M.}~\bibnamefont
  {Agostini}} \emph {et~al.} (\bibinfo {collaboration} {BOREXINO}),\ }\bibfield
   {title} {\enquote {\bibinfo {title} {{Comprehensive measurement of pp-chain
  solar neutrinos with Borexino}},}\ }\href {\doibase 10.22323/1.364.0400}
  {\bibfield  {journal} {\bibinfo  {journal} {PoS}\ }\textbf {\bibinfo {volume}
  {EPS-HEP2019}},\ \bibinfo {pages} {400} (\bibinfo {year} {2020})}\BibitemShut
  {NoStop}%
\bibitem [{\citenamefont {Bo}\ \emph {et~al.}(2024)\citenamefont {Bo} \emph
  {et~al.}}]{pandaX_B8}%
  \BibitemOpen
  \bibfield  {author} {\bibinfo {author} {\bibfnamefont {Z.}~\bibnamefont {Bo}}
  \emph {et~al.} (\bibinfo {collaboration} {PandaX}),\ }\bibfield  {title}
  {\enquote {\bibinfo {title} {{First Indication of Solar $^8$B Neutrino Flux
  through Coherent Elastic Neutrino-Nucleus Scattering in PandaX-4T}},}\ }\href
  {\doibase 10.1103/PhysRevLett.133.191001} {\bibfield  {journal} {\bibinfo
  {journal} {Phys. Rev. Lett.}\ }\textbf {\bibinfo {volume} {133}},\ \bibinfo
  {pages} {191001} (\bibinfo {year} {2024})},\ \Eprint
  {http://arxiv.org/abs/2407.10892} {arXiv:2407.10892 [hep-ex]} \BibitemShut
  {NoStop}%
\bibitem [{\citenamefont {O'Hare}(2021)}]{neutrino_fog}%
  \BibitemOpen
  \bibfield  {author} {\bibinfo {author} {\bibfnamefont {Ciaran A.~J.}\
  \bibnamefont {O'Hare}},\ }\bibfield  {title} {\enquote {\bibinfo {title}
  {{New Definition of the Neutrino Floor for Direct Dark Matter Searches}},}\
  }\href {\doibase 10.1103/PhysRevLett.127.251802} {\bibfield  {journal}
  {\bibinfo  {journal} {Phys. Rev. Lett.}\ }\textbf {\bibinfo {volume} {127}},\
  \bibinfo {pages} {251802} (\bibinfo {year} {2021})}\BibitemShut {NoStop}%
\end{thebibliography}%

\end{document}